\documentclass[5p,times]{elsarticle}

\usepackage{amssymb}
\usepackage{amsmath}
\usepackage{url}
\usepackage[final]{listings}
\usepackage{xcolor}
\usepackage{nth}

\definecolor{commentgreen}{RGB}{2,112,10}
\definecolor{eminence}{RGB}{108,48,130}
\definecolor{weborange}{RGB}{255,165,0}
\definecolor{frenchplum}{RGB}{129,20,83}

\lstnewenvironment{Python}
{\lstset {
language=Python,
frame=tb,
tabsize=4,
showstringspaces=false,
numbers=right,
breaklines=true,
commentstyle=\color{commentgreen},
keywordstyle=\color{eminence},
stringstyle=\color{blue},
basicstyle=\footnotesize\ttfamily, % basic font setting
emph={int,char,double,float,unsigned,void,bool},
emphstyle={\color{red}},
escapechar=\&,
classoffset=1, % starting new class
otherkeywords={>,<,.,;,-,!,=,~},
morekeywords={>,<,.,;,-,!,=,~},
keywordstyle=\color{orange},
classoffset=0,
numberstyle=\scriptsize,numbersep=2.0pt
}}
{}

\journal{Future Generation Computer Systems}

\usepackage{subcaption}

\begin{document}
\sloppy
\begin{frontmatter}



\title{Reduced and mixed precision turbulent flow simulations using explicit finite difference schemes}


\author[Pazmany,SOTON]{Bálint Siklósi\corref{cor1}} %% Author name
\ead{siklosi.balint@itk.ppke.hu}
\cortext[cor1]{Corresponding author.}

\author[SOTON]{Pushpender K. Sharma}
\author[JAXA]{David J. Lusher}
\author[Pazmany]{István Z. Reguly}
\author[SOTON]{Neil D. Sandham}

\affiliation[Pazmany]{organization={Pázmány Péter Catholic University },
            addressline={Szentkirályi u. 28}, 
            city={Budapest},
            postcode={1088}, 
            country={Hungary}}

\affiliation[SOTON]{organization={University of Southampton},
            addressline={Boldrewood Campus, Burgess Rd}, 
            city={Southampton},
            postcode={SO16 7QF}, 
            country={United Kingdom}}
\affiliation[JAXA]{organization={Japan Aerospace Exploration Agency (JAXA)},
            addressline={ 7-44-1 Jindaiji Higashi-machi, Chofu-shi}, 
            city={Tokyo},
            postcode={182-8522}, 
            country={Japan}}

%% Abstract
\begin{abstract}
The use of reduced and mixed precision computing has gained increasing attention in high-performance computing (HPC) as a means to improve computational efficiency, particularly on modern hardware architectures like GPUs. In this work, we explore the application of mixed precision arithmetic in compressible turbulent flow simulations using explicit finite difference schemes. We extend the OPS and OpenSBLI frameworks to support customizable precision levels, enabling fine-grained control over precision allocation for different computational tasks. Through a series of numerical experiments on the Taylor-Green vortex benchmark, we demonstrate that mixed precision strategies, such as half-single and single-double combinations, can offer significant performance gains without compromising numerical accuracy. However, pure half-precision computations result in unacceptable accuracy loss, underscoring the need for careful precision selection. Our results show that mixed precision configurations can reduce memory usage and communication overhead, leading to notable speedups, particularly on multi-CPU and multi-GPU systems.
\end{abstract}


%\item This study applies mixed precision in turbulent flow simulations.
%\item OPS and OpenSBLI now support customizable precision levels.
%\item Most configurations maintain accuracy using the Taylor-Green vortex benchmark.
%\item Numerical schemes show consistent performance across different precisions.
%\item Significant memory savings and speedups achieved with mixed precision.

%% Keywords
\begin{keyword}
Floating point precision \sep Mixed precision \sep Computational fluid dynamics\sep Computational performance and scalability \sep OpenSBLI\sep OPS\sep GPU


\end{keyword}

\end{frontmatter}

% \linenumbers


\section{Introduction}
\label{sec:intro}

The advent of artificial intelligence has led to the widespread adoption of reduced precision formats in hardware, such as half-precision (16-bit) floating-point arithmetic. While this transition presents opportunities for significant performance gains, it also raises concerns regarding the accuracy and stability of numerical results, particularly in high-performance computing (HPC) applications that are sensitive to precision. The challenge lies in determining the appropriate balance between precision and performance, as the demand for computational efficiency continues to grow.

Existing mixed precision techniques have shown promise in various applications. For instance, many linear solvers have been successfully implemented using mixed precision, allowing for improved performance without sacrificing accuracy~\cite{4531732,8948697,10.1145/1731022.1731027,Higham_Mary_2022,LUSZCZEK2024359}. Abdelfattah et al.~\cite{osti_1814447} reported advancements in mixed precision algorithms, including speedups in dense and sparse LU factorization, eigenvalue solvers, and GMRES implementations. FluidX3D~\cite{PhysRevE.106.015308}, a lattice Boltzmann CFD software, has explored mixed precision techniques, although it is limited to specific models and lacks flexibility in accommodating other simulation types. OpenFOAM~\cite{MixedOpenfoam}, another popular CFD framework, with its recent update, compiles its code in single precision while employing double precision exclusively for its linear algebra solvers. This approach allows OpenFOAM to leverage the efficiency of single precision across a wide range of simulations, although it primarily focuses on finite volume solvers.

Many finite difference methods (FDMs) are characterized by a common structural pattern: the iterative addition of a smaller value to a larger one at each computational step. This inherent structure presents an opportunity for strategic precision allocation that is exploited in the present contribution. In this approach, the larger value, which accumulates the results of multiple iterations and is more susceptible to accumulating roundoff errors, is typically maintained in a full precision format to minimize the loss of precision. Conversely, the smaller update values, which are discarded after each step and do not contribute to long-term error accumulation, can be represented in reduced precision formats, allowing for a controlled trade-off between precision and computational efficiency. By employing the OpenSBLI~\cite{OpenSBLI_LJS2021} framework, which automates the derivation of finite difference solvers, in conjunction with the OPS~\cite{Reguly_et_al_2018} library—designed for structured mesh solvers—we demonstrate the capability to generate optimized codes for multi-GPU and multi-CPU environments. OpenSBLI facilitates the generation of optimized reduced/mixed precision implementations of OPS codes, enabling efficient computations across various hardware configurations.

The flexibility afforded by OpenSBLI allows users to customize the precision of datasets utilized in their simulations. This capability is particularly advantageous in scenarios where computational efficiency is paramount, necessitating a careful balance between precision and performance. In this study, we present a series of tests conducted on a compressible Taylor-Green vortex simulation, which serves as a benchmark for evaluating the performance and accuracy of reduced precision computing. Our findings reveal that employing half, single, and double precision formats, as well as mixtures such as half-single and single-double, yields significant speedups without compromising the numerical accuracy of the results. Notably, only the half-precision runs exhibited numerically unsatisfactory outcomes, highlighting the critical importance of precision selection in computational simulations.

The main contributions of this work are as follows:

\begin{itemize}
    \item Extended OPS for Mixed Precision: We have enhanced the OPS library to support mixed precision operations, allowing for improved computational efficiency in structured mesh solvers.
    \item Automated Mixed Precision Implementation in OpenSBLI: We extended the OpenSBLI framework to automate the generation of mixed precision implementations, simplifying the integration of these strategies into existing workflows.
    \item Proposed Mixed Precision Algorithm: Our study introduces a mixed precision algorithm that optimally maintains work and residual arrays in lower precision, effectively balancing memory usage and computational accuracy.
    \item Accuracy Validation with Taylor-Green Vortex: We demonstrated that all configurations, except for pure half precision (HP), maintain acceptable accuracy levels when evaluated using the Taylor-Green vortex benchmark.  Importantly, we found that the accuracy of the results is independent of various simulation parameters, such as Reynolds number and Mach number. Additionally, multiple numerical schemes are affected similarly across different precision levels; thus, while choosing the appropriate scheme remains a critical factor for achieving accurate results, its performance is consistent regardless of whether reduced precision is employed.
    \item Comprehensive Performance Analysis: We provided a detailed performance evaluation focusing on memory efficiency and communication overhead, showcasing significant speedups achieved through our mixed precision strategies without compromising result integrity.

\end{itemize}

The paper is structured as follows: Section \ref{sec:matmet} provides a detailed discussion of floating-point arithmetic, including an introduction to half precision and the potential risks associated with using reduced precision, such as loss of accuracy and numerical instability. We also introduce the OPS and OpenSBLI frameworks, outlining their relevance to high-performance computing applications. Section \ref{sec:mixdopensbli} focuses on the algorithmic changes required to implement mixed precision strategies effectively within finite difference methods. Section \ref{sec:results} presents our testing and evaluation, starting with an introduction to the Taylor-Green vortex test case. We then assess the accuracy of various precision configurations and subsequently analyze the performance, highlighting both the computational speedups and memory savings achieved with mixed precision. Finally, in Section \ref{sec:concl}, we discuss conclusions and future work, highlighting the potential of mixed precision techniques in larger and more complex applications, as well as improvements in half-precision performance and flexibility in the OPS framework.

\section{Methodology / Background}
\label{sec:matmet}

\subsection{Floating-Point Arithmetic in CFD Applications}

Floating-point arithmetic is a critical component in computational fluid dynamics (CFD) applications, enabling the representation and manipulation of real numbers in a way that can accommodate a wide range of values. This section provides an overview of floating-point arithmetic, its operational mechanics, the risks associated with using insufficient precision, and a discussion on half precision.
\subsubsection{Overview of Floating-Point Arithmetic}
Floating-point numbers are represented in computers using a format that consists of three main components: the sign bit, the exponent, and the significand (or mantissa). This representation allows for the encoding of very large and very small numbers by scaling the significand with the exponent. The standard format for floating-point representation is defined by the IEEE 754 standard~\cite{IEEE754}, which specifies several precision levels, including single precision (32 bits) and double precision (64 bits).
In a typical floating-point representation, a number is expressed as: $value=(-1)^{sign} \times significand \times 2^{exponent}$. This format allows for efficient arithmetic operations such as addition, subtraction, multiplication, and division, which are fundamental in CFD simulations that require iterative calculations over vast datasets.

\subsubsection{Dangers of Using Too Small Precision}
Using insufficient precision in floating-point arithmetic can lead to significant computational errors, particularly in CFD applications where numerical stability is crucial. Key issues include:
\begin{itemize}
    \item Round-off Errors: When calculations exceed the precision limit (caused by the fixed length of the mantissa), the results can become inaccurate due to rounding. This is particularly problematic in iterative methods where errors can accumulate and lead to divergent results. For examples, see \cite{10.1007/3-540-47764-0_14}.
    \item Overflow and Underflow: In floating-point arithmetic, numbers can exceed the maximum representable value (overflow) or fall below the minimum in absolute value (underflow). Both scenarios can lead to catastrophic failures in simulations, causing incorrect physical predictions.
    \item Loss of Significance: After converting to floating-point and subtracting two nearly equal numbers, significant digits can be lost, resulting in a substantial relative error. This is known as catastrophic cancellation, which can severely affect the accuracy of CFD simulations. For examples, see \cite{10.1145/2345156.2254118}
\end{itemize}
These issues underscore the importance of selecting an appropriate precision level that balances computational efficiency with the accuracy required for the specific CFD application.

\subsubsection{Half Precision Floating-Point Arithmetic}
Half precision floating-point arithmetic (FP16) has recently gained significant traction in high-performance computing (HPC). The FP16 format, which utilizes 16 bits for numerical representation, strikes a balance between computational efficiency and memory usage, making it particularly well-suited for modern GPU architectures. Recent advancements in NVIDIA GPUs, such as the Volta and Ampere architectures, have introduced native support for FP16, enabling substantial performance improvements in various scientific computations.
The use of FP16 can lead to remarkable speedups in computational tasks. For instance, one study~\cite{8665777} has demonstrated that employing mixed precision techniques, which leverage FP16 alongside higher precision formats like FP32 or FP64, can achieve speedups of up to 4 times in iterative refinement solvers. This is primarily due to the enhanced computational throughput provided by Tensor Cores in NVIDIA GPUs, which are optimized for half precision arithmetic. Additionally, research has shown that FP16 can significantly reduce memory bandwidth requirements, allowing for faster data transfers and improved overall performance in large-scale simulations~\cite{Wan2024EnhancingCE}.
However, the transition to half precision is not without challenges. The limited range and precision of FP16 can introduce numerical instability, particularly in operations that require high accuracy. Specifically, catastrophic cancellation can result in substantial relative errors, particularly in time-marching methods where small perturbations can accumulate. 

To address these challenges, researchers have proposed various strategies, such as iterative refinement and scaling techniques, to enhance the accuracy of computations while leveraging the performance benefits of FP16. For example, recent work has shown that using a modified LU factorization can produce more numerically stable results, effectively mitigating the risks associated with limited precision formats~\cite{8916392}. These approaches illustrate that while FP16 offers significant advantages in terms of speed and efficiency, careful consideration of numerical stability is essential to maintain the integrity of computational results in FDM CFD applications.
In addition to half precision (FP16), another format gaining attention is bfloat16~\cite{intel_bfloat16}, which retains the exponent size of FP32 while reducing the significand to 7 bits. This design allows bfloat16 to maintain a similar dynamic range as FP32, making it particularly useful in deep learning applications where training stability is crucial. However, despite its advantages, bfloat16 still has limitations in terms of precision, which can lead to numerical instability in certain computational scenarios. Furthermore, there are even smaller floating-point formats in use, such as float8~\cite{TORTORELLA2023122} and other reduced precision types, but due to the challenges and inaccuracies encountered with FP16, we do not advocate for going lower in precision.

\subsection{OPS Domain-Specific Library}

The Oxford Parallel library for Structured mesh solvers (OPS)~\cite{Reguly_et_al_2018} is a domain-specific library designed to streamline the implementation of high-performance applications in structured mesh computations. OPS provides a high-level abstraction for stencil operations commonly used in numerical methods, allowing developers to focus on algorithm development without the complexities associated with low-level parallel programming. This abstraction significantly enhances the ease of use and maintainability of the code, which is particularly beneficial for researchers and practitioners in the field.
A notable feature of OPS is its capability to handle multiple datasets, each of which can be assigned its own precision. This flexibility allows for efficient management of different numerical requirements within a single application, accommodating varying precision needs across datasets. By enabling developers to specify precision at the dataset level, OPS facilitates the creation of applications that can dynamically adapt to the precision requirements of different computations, optimizing performance while ensuring accuracy.
OPS is designed with performance portability in mind, allowing applications to run efficiently across a range of hardware architectures, including CPUs and GPUs. This portability is achieved through a combination of automated code generation and optimization techniques that adapt to the specific features of the underlying hardware, ensuring that applications can leverage the full capabilities of diverse computing environments. Furthermore, OPS supports various programming models, including OpenMP, MPI, and CUDA, providing developers with the flexibility to choose the most suitable approach for their specific application and hardware configuration.
Overall, OPS represents a powerful tool for simplifying the development of structured mesh applications. Its focus on ease of use, performance portability, and support for multiple programming models makes it an attractive option for researchers looking to implement efficient and scalable numerical methods in their computational workflows.

\subsection{OpenSBLI}

OpenSBLI \citep{OpenSBLI_LJS2021} is a complete code generation system for computational fluid dynamics that automatically generates OPS code starting from a compact description of the governing equations using subscript notation. It uses symbolic Python to expand the governing equations and carry out the subsequent discretisation, using high-order finite differences. The front end is used to define and expand the governing equations and constituent relations. It also defines all the boundary and initial conditions and sets all the run time parameters. This provides an effective implementation of the `separation of concerns' work flow \citep{Ober2017}, whereby users focusing on the fluid dynamics can carry out most tasks from the python front end. Alternatively, numerical methods developers can work either in the OPSC code or in the OpenSBLI code generation, while computer-science-based performance optimisations can be carried out in the OPS translator. The main applications of OpenSBLI are direct numerical simulation (DNS) and large eddy simulation (LES) of compressible-flow problems involving transition to turbulence or fully-developed turbulence on structured grids.

An initial version (V1) of the software \citep{JACOBS201712} demonstrated the code-generation concept for simple low-speed periodic flows and the benefits of the OPS code translation in being able to run efficiently on heterogeneous computing architectures. The project was restarted from a clean code base in \cite{LJS2018}, which led to the subsequent public code release in \cite{OpenSBLI_LJS2021}. This version (V2) included shock capturing using Weighted Essentially Non-Oscillatory (WENO) and less dissipative Targeted Essentially Non-Oscillatory (TENO) schemes along with a greatly expanded set of boundary-conditions and application demonstrations, including shock-wave/boundary-layer interactions \citep{LJS2018} and channel flow validation test cases \citep{LC2022_TCF_HS}. This release also included generalised curvilinear geometries. The current release version of OpenSBLI is V3 \cite{OpenSBLI_2024_CPC_V3}. Version 3 added multi-block mesh support, airfoil simulations, new numerical methods and filters, and various performance and efficiency improvements. Reduced-dimension I/O, partial reductions for spanwise averaging, and mixed-precision support were also added, via upgrades to both OpenSBLI and the OPS library. The combination of multi-block capability and new filter-based shock-capturing schemes \citep{OpenSBLI_2024_CPC_V3} has led to new applications of implicit LES simulations of flow over airfoils, including state-of-the-art studies of airfoil buffet in the transonic flow regime \citep{LSH2024_narrow_buffet,lusher2025highfidelity_JFM} that exploit large GPU-based machines to perform high-fidelity simulations on the order of $N\sim 10^{10}$ mesh points.

\subsubsection{Numerical Methods and Split Formulations of the Equations}
The accuracy of low- and mixed-precision algorithms is assessed in this work in the context of the unsteady full-3D Navier-Stokes equations for a compressible Taylor-Green vortex problem described in Section~\ref{sec:TGV_description}. All simulations are performed using explicit \nth{4}-order accurate non-dissipative central-differencing. Both the convective and diffusive parts of the Navier-Stokes equations are solved at \nth{4}-order to maintain consistent spatial order throughout. Time-stepping is performed by a low-storage explicit \nth{3}-order Runge-Kutta scheme. Details of the implementation have previously been given in \cite{OpenSBLI_LJS2021}.

Depending partly on the magnitude of the Reynolds number, direct application of standard central derivative approximations to the Navier-Stokes equations can lead to numerical instabilities \citep{Pirozzoli_review}. This is due to an accumulation of aliasing errors that occurs from discrete evaluation of the product between two or more terms within the non-linear convective derivatives. The lack of numerical robustness can also be linked to the failure of standard formulations to discretely preserve quadratic invariants such as global kinetic energy in the inviscid limit \citep{Coppola_ASME2019}. To alleviate these discretisation issues, convective terms of the base Navier-Stokes equations are routinely reformulated in modern CFD codes in what are known as split formulations. These alternative formulations of the governing equations have been reported to improve numerical robustness via reduced aliasing errors and preservation of certain invariant quantities \citep{Coppola_CubicSplit_2019}.

Various split-forms are available in the literature \citep{Coppola_CubicSplit_2019}. They typically focus on the reformulation of non-linear convective derivative terms that take the general form
\begin{equation}\label{eq:convective_form}
\mathcal{C}=\frac{\partial \rho u_j \varphi}{\partial x_j},
\end{equation}
where $\varphi$ takes the value of $\left(1, u_i, E\right)$, for the continuity, momentum, and energy components of the Navier-Stokes equations, respectively. As an example of one of the split-forms available in OpenSBLI, the Feiereisen quadratic split-form \cite{feiereisen1981numerical}, expands these terms quadratically as
\begin{equation}
\frac{\partial \rho u_j \varphi}{\partial x_j} \rightarrow \frac{1}{2} \frac{\partial \rho u_j \varphi}{\partial x_j}+\frac{1}{2}\left(\varphi \frac{\partial \rho u_j}{\partial x_j}+\rho u_j \frac{\partial \varphi}{\partial x_j}\right).
\end{equation}
While the underlying physical equations are mathematically equivalent in an algebraic sense, the split-form is considerably more robust numerically when the equations are evaluated on a discrete mesh with finite-difference approximations \citep{Coppola_CubicSplit_2019}. The split-forms are computationally more expensive due to the increased number of floating-point operations required, however, as CFD codes are typically memory-bound, they are an efficient way of improving numerical stability for large-scale calculations at high Reynolds numbers. In this work, we apply the quadratic Blaisdell split-form \citep{BLAISDELL1996207} of the equations to improve numerical robustness throughout. As we are interested in the effect of low- and mixed-precision and the propagation of numerical errors in finite-difference approximations, other quadratic and cubic split-forms are also tested in the inviscid limit with varying numerical precision in Section~\ref{sec:accuracy}. 

\section{Enabling mixed precision in OpenSBLI}
\label{sec:mixdopensbli}

\subsection{Mixed precision using explicit finite difference methods}\label{sec:algorithm}

For simulation of transitional and turbulent compressible flows, in which a wide range of spatial and temporal scales are present, explicit finite difference schemes are commonly used. High-order finite differences are used for evaluating spatial derivatives, while low-storage variants of Runge-Kutta time advance schemes are adopted. The conservation form of the governing Navier-Stokes equations is used, with the vector of flow variables given by ${\bf{Q}}=(\rho, \rho u, \rho v, \rho w, \rho E)^T$, where $\rho$ is the density, $u$, $v$, and $w$ are the velocity components and $E=e+(u^2+v^2+w^2)/2$ is the total energy per unit mass, adding the internal energy per unit mass $e$ to the kinetic energy per unit mass. The complete governing equations are given in \citep{Lusher2020_TGV}. Here, we focus on the time advance step to explain how reduced precision schemes may be implemented.

\begin{figure}[t]    
    \centering
    \includegraphics[trim=0cm 2cm 0cm 2cm,clip,width=.9\linewidth]{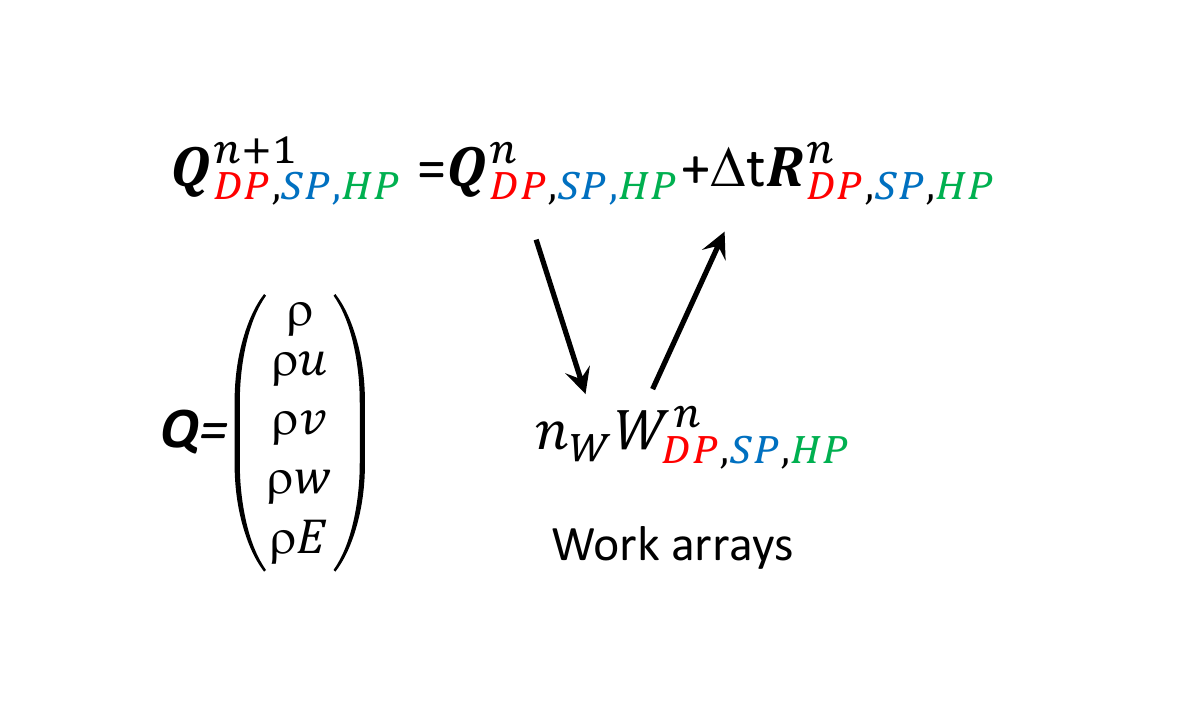}
    \caption{Schematic of the use of $n_W$ variable precision work arrays $W$ to form residuals $R$ for update of conservative variables $Q$ during a typical Runge-Kutta substep (DP=double precision, SP=single precision, HP=half precision).}
    \label{fig:Storage_schematic}
\end{figure}

The conservative flow variables are advanced in time using compact Runge-Kutta methods based on storage of $\bf Q$ and a change denoted $\bf \tilde Q$. At each substep $i$ these storage locations are updated according to
\begin{equation}
    {\bf \tilde Q}^i=A_i {\bf \tilde Q}^{i-1}+\Delta t {\bf R}^{i-1}
    \label{eq:RKmethod1}
\end{equation}
and
\begin{equation}
    {\bf Q}^i={\bf Q}^{i-1}+B_i{\bf \tilde Q}^i,
\end{equation}
where $A_i$ and $B_i$ are scalar coefficients of the scheme, $\Delta t$ is the time step and $\bf R$ is the residual, containing all the remaining terms from the governing equations. After $m$ substeps the solution at the next time level $n+1$ is given by
\begin{equation}
    {\bf Q}^{n+1}={\bf Q}^m.
\end{equation}
The update procedure is shown schematically in Figure \ref{fig:Storage_schematic} based on a simple Euler update, but containing the essential features of equation \ref{eq:RKmethod1}. The residual $\bf R$ needs to be computed from the solution $\bf Q$ at the previous step. This step contains most of the computational cost of the algorithm and is typically accomplished using a number $n_W$ of work arrays $W$, where $n_W$ may be of the order of 20, but can be much more if curvilinear co-ordinates are used. The work arrays typically are the datasets containing first and second derivatives of various quantities which are eventually used to evaluate the residual, $\bf R$.
% Schematic

To discuss the use of mixed precision algorithms, we first note that each of the array types $\bf Q$ (flow variables), $\bf R$ (residuals) and $W$ work arrays could in principle be represented using different numerical precision. A standard treatment would be to store all in double precision (denoted DP and shown in red in figure \ref{fig:Storage_schematic}). One could in principle also do all the operations in single precision (SP, blue) or even half precision (HP, green). In this paper, we will also consider mixed precision cases; for example, a single/double mixed precision case (SPDP) can be defined where we retain $\bf Q$ in double precision, while using single precision for $\bf R$ and all the $W$. Similarly, a half/single mixed precision case (HPSP) stores $\bf Q$ in single precision and $\bf R$ and $W$ in half precision. Clearly, there are other options that we will discuss later. One could also split the residual $\bf R$ into different components that could be treated with different precision, as proposed in~\cite{Schlatter2024Impact}.

Two approaches to the work array usage are applied in this study. The first one is called the ``default" method, and the second is called the ``Storesome" method. The default method in OpenSBLI creates various 3D work arrays to calculate and store the residual terms in memory before passing them to the main kernel that does the final evaluation of $\bf R$ involving the inviscid and viscous fluxes. This method is memory intensive, but helps with code structure and readability, mimicking what was previously done with manually-written code. The second method, called the ``Storesome" method, only uses a few of the 3D work arrays and computes most of the derivatives directly within the kernel functions, as and when required. Jammy et al \cite{Jammy2019} demonstrated significant improvements with the Storesome approach, in terms of memory usage and run time.

\subsection{Generation of mixed-precision source codes using OpenSBLI}
The most recent version of OpenSBLI \citep{OpenSBLI_2024_CPC_V3} was extended for this work to implement the different low- and mixed-precision algorithm strategies. This required the following changes to be made to the internal code-generation engine:

\begin{itemize}
    \item A new user interface option in the high-level Python script for specification of the precision to be used globally within the simulation (double/single/half).
    \item Modifications to the array declaration types and input/output arguments to OPS parallel loops based on the selection made by the user.
    \item Explicit C/C++ casting of quantities appearing on the right-hand-side of equations in the simulation code, as required.
    \item A Python interface for specifying a lower/higher precision for certain array quantities (e.g. $\bf Q$, $\bf R$, $\bf W$ discussed in Section~\ref{sec:algorithm}) to enable mixed-precision strategies.
\end{itemize}

\noindent The new interface for specifying lower- and mixed-precision is shown in the example code snippet below:

% \begin{Python}[caption={Selecting mixed-precision inputs in OpenSBLI V3 \citep{OpenSBLI_2024_CPC_V3}},captionpos=t]
\begin{Python}
# Set the global simulation precision to single
SimulationDataType.set_datatype(FloatC)
# Define custom mixed-precision options using presets
mixed_precision_config = {
'q_vector' : ([], Double),
'RK_arrays' : ([], Double),
'residuals' : ([], FloatC),
'wk_arrays' : ([], FloatC),
'casting' : 'explicit'}
# Optionally, select custom arrays to set the precision of
custom_arrays = [block.location_dataset(dset) for dset in ['T', 'p']]
mixed_precision_config['custom'] = (custom_arrays, Half)
# Call the OPS C code-generator 
OPSC(alg, mixed_precision_config)
\end{Python}

In line 2, the default global precision of the simulation is specified by the user as single (\texttt{FloatC}). A configuration dictionary is then created in lines 4-9 to specify mixed-precision alterations to the global precision depending on the storage quantity in question. Four presets are shown between lines 5-8, which, in this example, raise the precision of the conservative ${\bf Q}$ and temporary Runge-Kutta ${\bf \tilde Q}$ storage arrays to double precision. Other available presets for residual $\bf R$ and work-arrays $\bf W$ are also shown. For completeness, a custom array option is also shown in lines 11-12. Here, the arrays for temperature and pressure ($T, p$) are set as half precision, to provide additional flexibility to explore more complex mixed-precision strategies. Finally, OpenSBLI's OPS C code-writer is called in line 14 with the requested mixed-precision configuration. Internally, the code-generation engine will define the precision of these quantities and their API calls within the OPS library automatically based on the user input. 

\section{Results}
\label{sec:results}

\subsection{Taylor Green test case}\label{sec:TGV_description}

The Taylor-Green vortex problem has become a standard test case for assessing the accuracy and efficiency of schemes to solve the unsteady Navier-Stokes equations, particularly of high-order schemes \citep{DeBonis2013}. It is a straightforward case to set up, consisting of a triply-periodic domain with a uniformly spaced grid and a prescribed analytic initial condition. From this starting condition, the flow evolves into discrete vortices which subsequently break down to a turbulent flow state. The kinetic energy of the initial condition eventually dissipates into heat and the accuracy of the method is assessed by monitoring the time evolution of the turbulence kinetic energy and the dissipation rate. Both incompressible and compressible forms of the problem can be specified. The effect of Mach number was demonstrated by Lusher \& Sandham \citep{Lusher2020_TGV} and one of their cases was used for a 7-way multi-code comparison with a variety of numerical methodologies in \cite{NATO2024_TGV}. 

The starting equations for the Taylor-Green vortex problem are given by

% \begin{eqnarray} \label{eq:initial_cond}
% \label{eq:eq_u} u(x,y,z,t=0) &=& \scalebox{1.1}{$\sin\left(x\right) \cos\left(y\right) \cos\left(z\right)$} , \\
% \label{eq:eq_v} v(x,y,z,t=0) &=& \scalebox{1.1}{$-\cos\left(x\right) \sin\left(y\right) \cos\left(z\right)$} , \\
% \label{eq:eq_w} w(x,y,z,t=0) &=& \scalebox{1.1}{$0$} , \\
% \label{eq:eq_p} p(x,y,z,t=0) &=& \scalebox{1.1}{$\frac{1}{\gamma M^2} + \frac{1}{16} \left[ \cos\left(2x\right) + \cos\left(2y\right) \right] \left[ 2+\cos\left(2z\right) \right]  .$}
% \end{eqnarray}

\begin{align} 
    \label{eq:eq_u} u(x,y,z,t=0) &= \sin(x) \cos(y) \cos(z), \\
    \label{eq:eq_v} v(x,y,z,t=0) &= -\cos(x) \sin(y) \cos(z), \\
    \label{eq:eq_w} w(x,y,z,t=0) &= 0, \\
    \label{eq:eq_p} p(x,y,z,t=0) &= \frac{1}{\gamma M^2} + \frac{1}{16} \left[ \cos(2x) + \cos(2y) \right] \left[ 2 + \cos(2z) \right].
\end{align}

Here, $M$ is the reference Mach number, while $\gamma = 1.4$ for air. The equations are solved in a non-dimensional form and all the quantities are non-dimensionalised with corresponding reference values. Also, the initial density $\rho$ at the start of the simulations is evaluated based on the non-dimensional form of the equation of state, i.e., 
\begin{equation}
    \label{eq:rho_ref} \rho(x,y,z,t=0) =   \gamma M^2 p.
\end{equation}
In the TGV problem, a constant initial reference temperature is assumed across the entire domain at the beginning at $t=0$, hence the non-dimensional temperature is equal to one and doesn't feature in Eq. \ref{eq:rho_ref}.

All the simulations are carried out with fourth order central differencing. The solution domain size can be either a $(2\pi)^3$ or (exploiting symmetries in the initial condition) $\pi^3$. Key parameters are the Reynolds and Mach numbers. For the tests of reduced precision, we adopt the triply symmetric case (denoted TGSym) that was part of the 2021 OpenSBLI release \citep{OpenSBLI_LJS2021} with a Reynolds number set to $Re=800$ to make the case better resolved on modest grids (compared to the $Re=1600$ used in \cite{NATO2024_TGV} requiring grids between $512^3$ and $2048^3$ for a supersonic case). A default resolution of $256^3$ is adopted here, which is enough for these cases to be fully resolved (equivalent to $512^3$ in a $(2\pi)^3$ domain). Using an intermediate Mach number of $M=0.5$ allows larger time steps and is more representative of compressible flow applications than the choice of $M=0.1$ that is commonly used to compare with incompressible simulations; however, we will also consider the effect of Mach number, since we wish to check for any issues in using reduced precision towards the incompressible limit. We also later consider an inviscid version of the problem to study the stability of various splitting schemes in the context of reduced precision algorithms.

\begin{figure*}[htbp]
    \centering
    \begin{subfigure}[b]{0.49\linewidth}
        \centering
        \includegraphics[width=\linewidth]
        {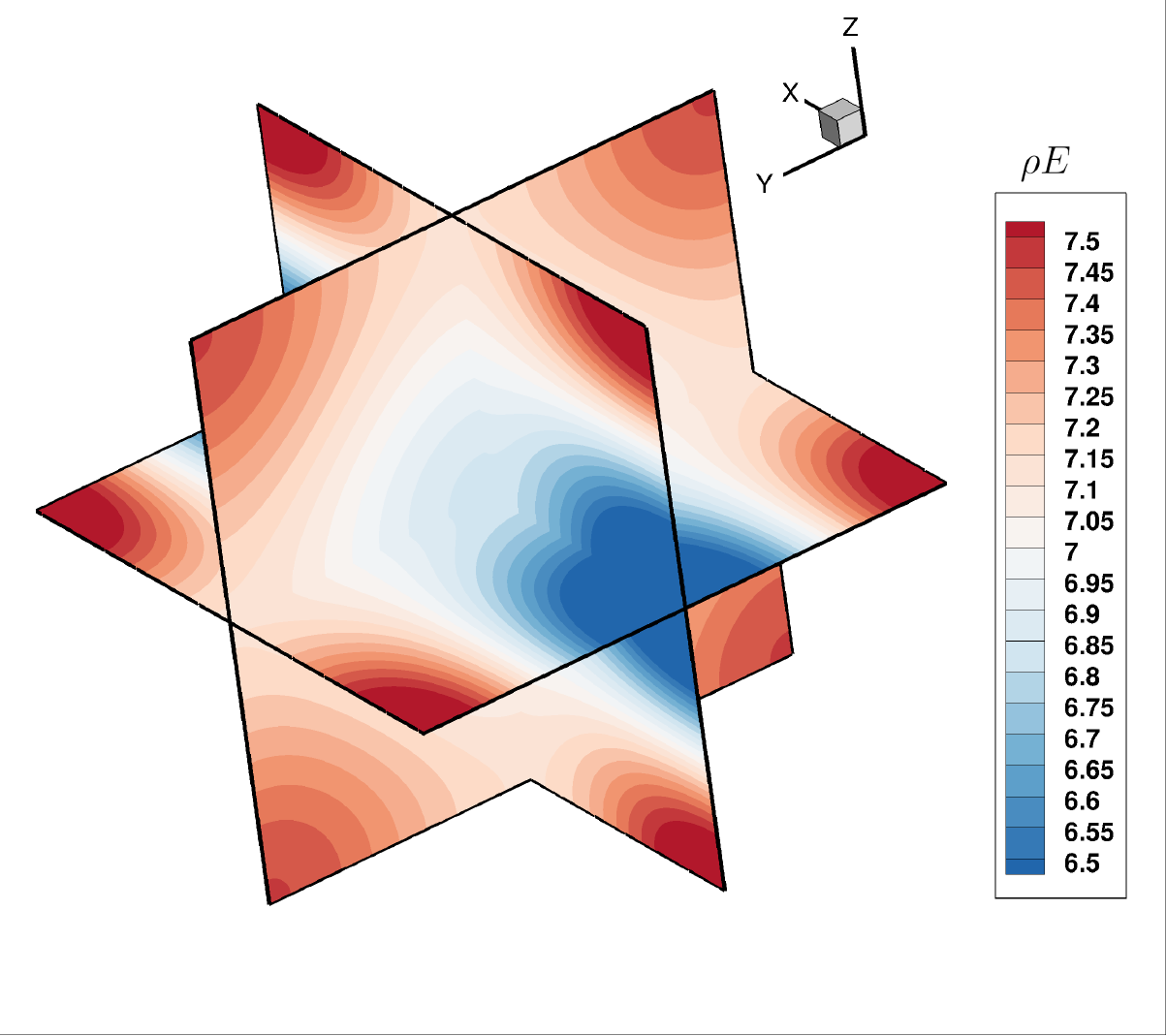}
        \caption{$t=0$}
        \label{fig:TGV_evolutions_t_0}
    \end{subfigure}
    \hfill
    \begin{subfigure}[b]{0.49\linewidth}
        \centering
        \includegraphics[width=\linewidth]
        {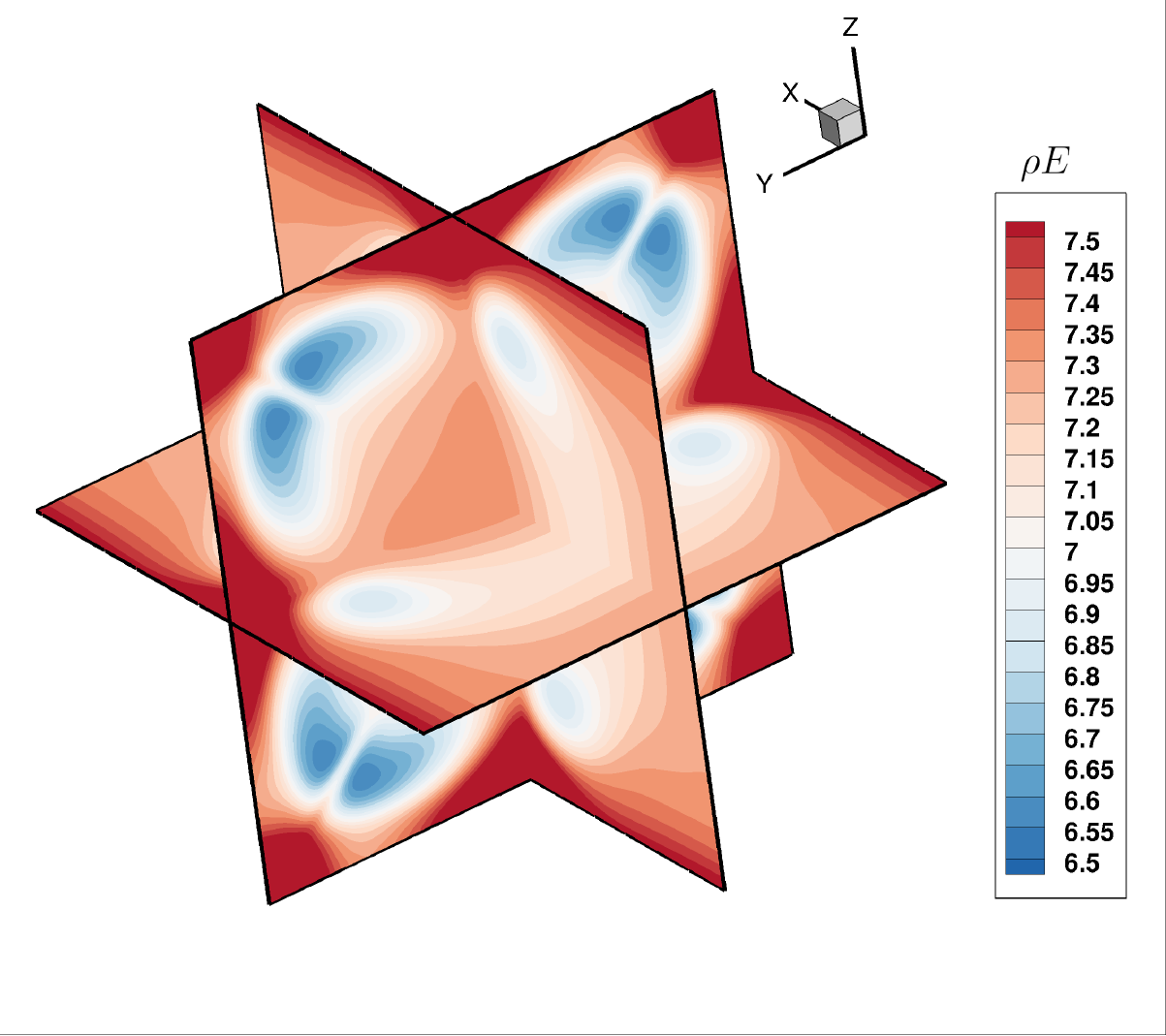}
        \caption{$t=5$}
        \label{fig:TGV_evolutions_t_5}
    \end{subfigure}
    \begin{subfigure}[b]{0.49\linewidth}
        \centering
        \includegraphics[width=\linewidth]
        {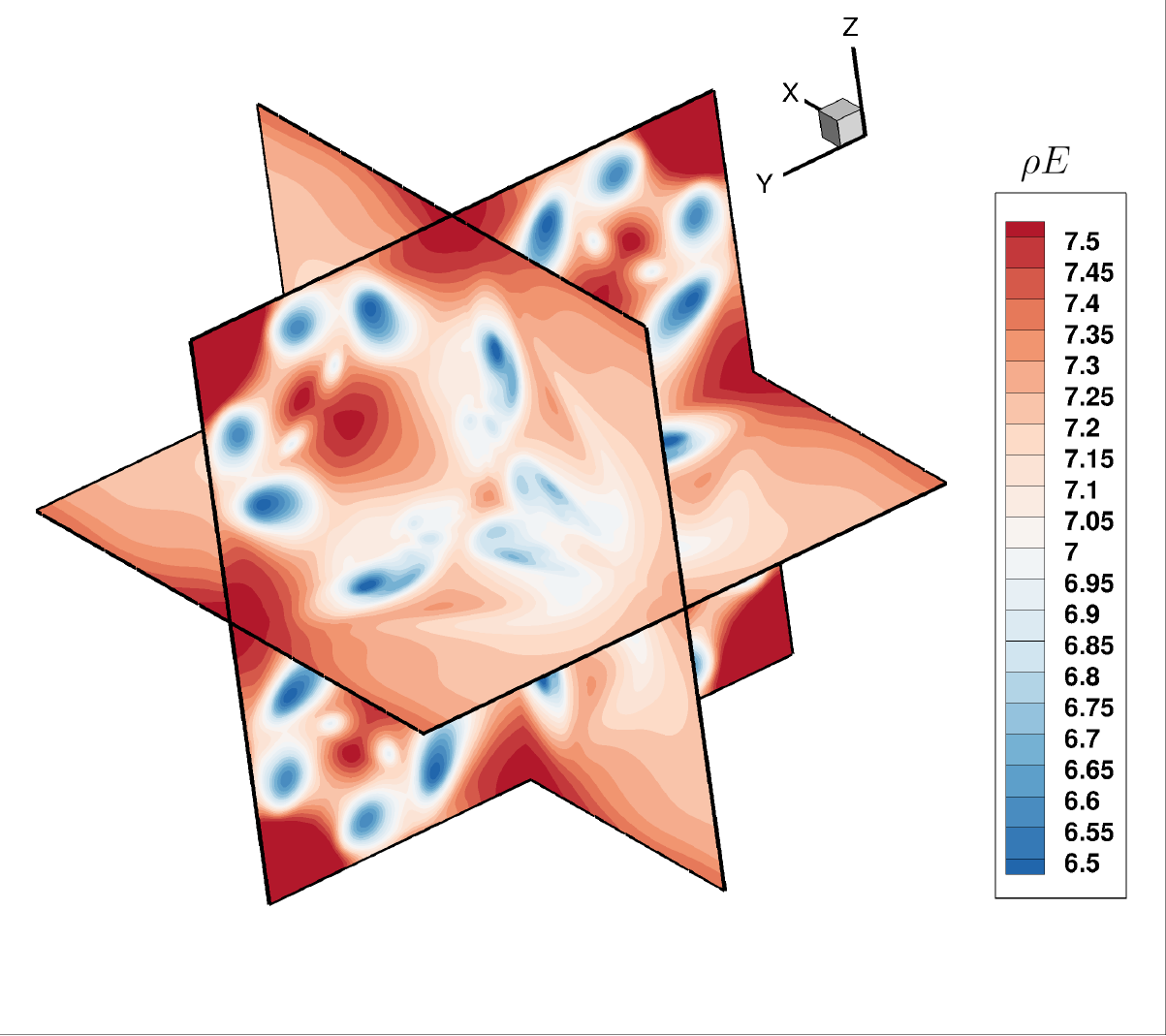}
        \caption{$t=10$}
        \label{fig:TGV_evolutions_t_10}
    \end{subfigure}
    \begin{subfigure}[b]{0.49\linewidth}
        \centering
        \includegraphics[width=\linewidth]
        {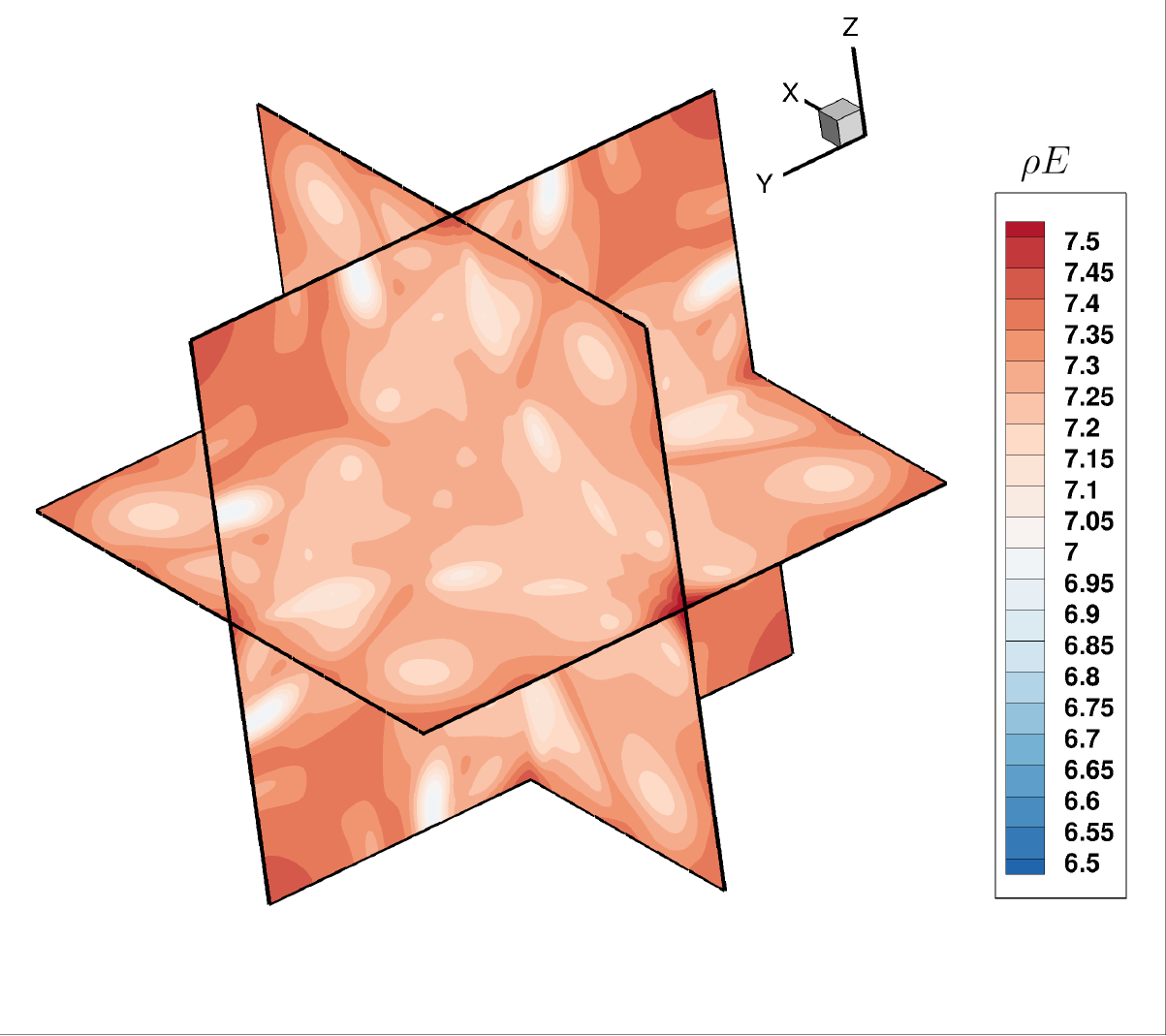}
        \caption{$t=15$}
        \label{fig:TGV_evolutions_t_15}
    \end{subfigure}    
    \caption{Contours of $\rho E$ in three mutualy perpendicular slices at the mid locations in $x$, $y$ and $z$-directions, demonstrating the evolution of TGV state at different times: (a) t=0, (b) t=5, (c) t=10 and (d) t=15.}
    \label{fig:TGV_evolutions}
\end{figure*}

Figure \ref{fig:TGV_evolutions} shows a typical evolution of the flow field at a few different time instances. Contours of $\rho E$ are shown for a $M = 0.5$ TGV simulation run in double precision. The first frame (Fig.~\ref{fig:TGV_evolutions_t_0}) shows the smooth state of the flow at time $t=0$, as per the description in Eqs. \ref{eq:eq_u}-\ref{eq:eq_p}. As time evolves, the evolution of the flow leads to the formation of smaller and smaller scale vortical structures (Figs.~\ref{fig:TGV_evolutions_t_5} and \ref{fig:TGV_evolutions_t_10}), illustrating the cascade process of turbulent flow. With no production of turbulence, the flow (Fig.~\ref{fig:TGV_evolutions_t_15}) slowly decays.

Before we move to quantitative results, we outline the physical quantities of interest in the TGV problem to assess the efficacy of various precision calculations presented in this research. The volume-averaged kinetic energy at each instance of time is defined as
\begin{equation}
    \label{eq:KE} K =    \frac{1}{ V } \int_V \frac{1}{2} u_i u_i dV.
\end{equation}
Here, $V$ is the volume of the computational box and is equal to $\pi^3$ for the symmetric case. We will also look at the evolution of solenoidal dissipation, which represents the enstrophy contribution to dissipation. The simulations performed in the present paper are all for subsonic Mach numbers where the dilatation part of dissipation is negligible. The equation for solenoidal dissipation is
\begin{equation}
    \label{eq:K} \epsilon^S = \frac{1}{ Re } \int_V \left(\epsilon_{ijk}\frac{\partial u_k}{\partial x_j} \right)^2 dV.
\end{equation}
Here, $Re$ is the simulation Reynolds number, while the integrand represents the inner product of the vorticity vector, also known as enstrophy.

\begin{figure}
    \centering
    \includegraphics[width=0.9\linewidth]{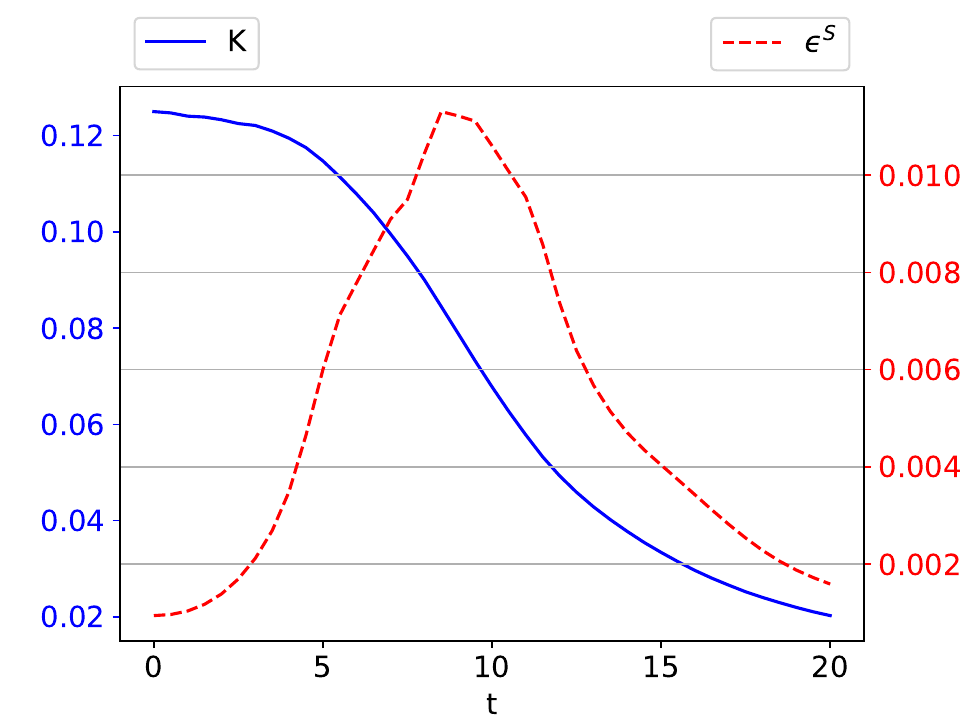}
    \caption{Kinetic energy (K) and dissipation ($\epsilon^S$) relative to the time.}
    \label{fig:KE-dissip}
\end{figure}

Figure \ref{fig:KE-dissip} shows the behavior of $K$ (left hand scale) and $\epsilon^S$ (right hand scale) over a simulation time from $t=0$ to $t=20$. For spatially homogeneous decaying turbulence, the rate of change of $K$ is proportional to $-\epsilon$, so the kinetic energy decays monotonically once small scale structures have formed and the dissipation, which is positive definite, becomes significant. Given the derivative nature of the relation between $K$ and $\epsilon$, the solenoidal dissipation rate $\epsilon^S$, representing almost all of $\epsilon$, is a more sensitive measure of the numerical accuracy that is used to compare the different techniques employed in this study.

\subsection{Accuracy}\label{sec:accuracy}

\begin{figure}[t]
    \centering
    \begin{subfigure}[b]{.8\linewidth}
        \centering
        \includegraphics[width=\linewidth]{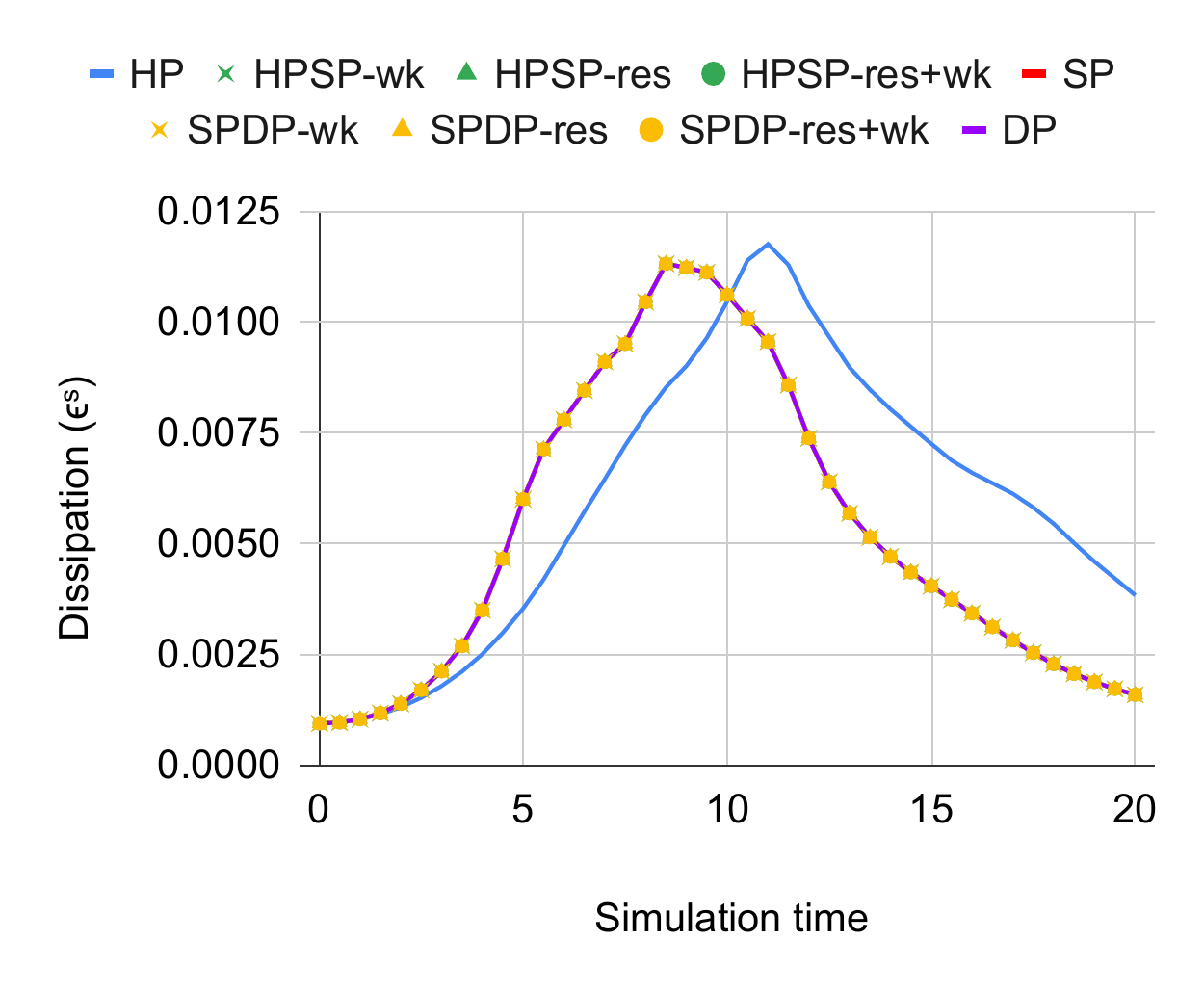}
        \caption{Dissipation\\}%\phantom{B}}

        \label{fig:TGsym_256c_dt0p0025_precision_diss}
    \end{subfigure}
    \hfill
    \begin{subfigure}[b]{.8\linewidth}
        \centering
        \includegraphics[width=\linewidth]{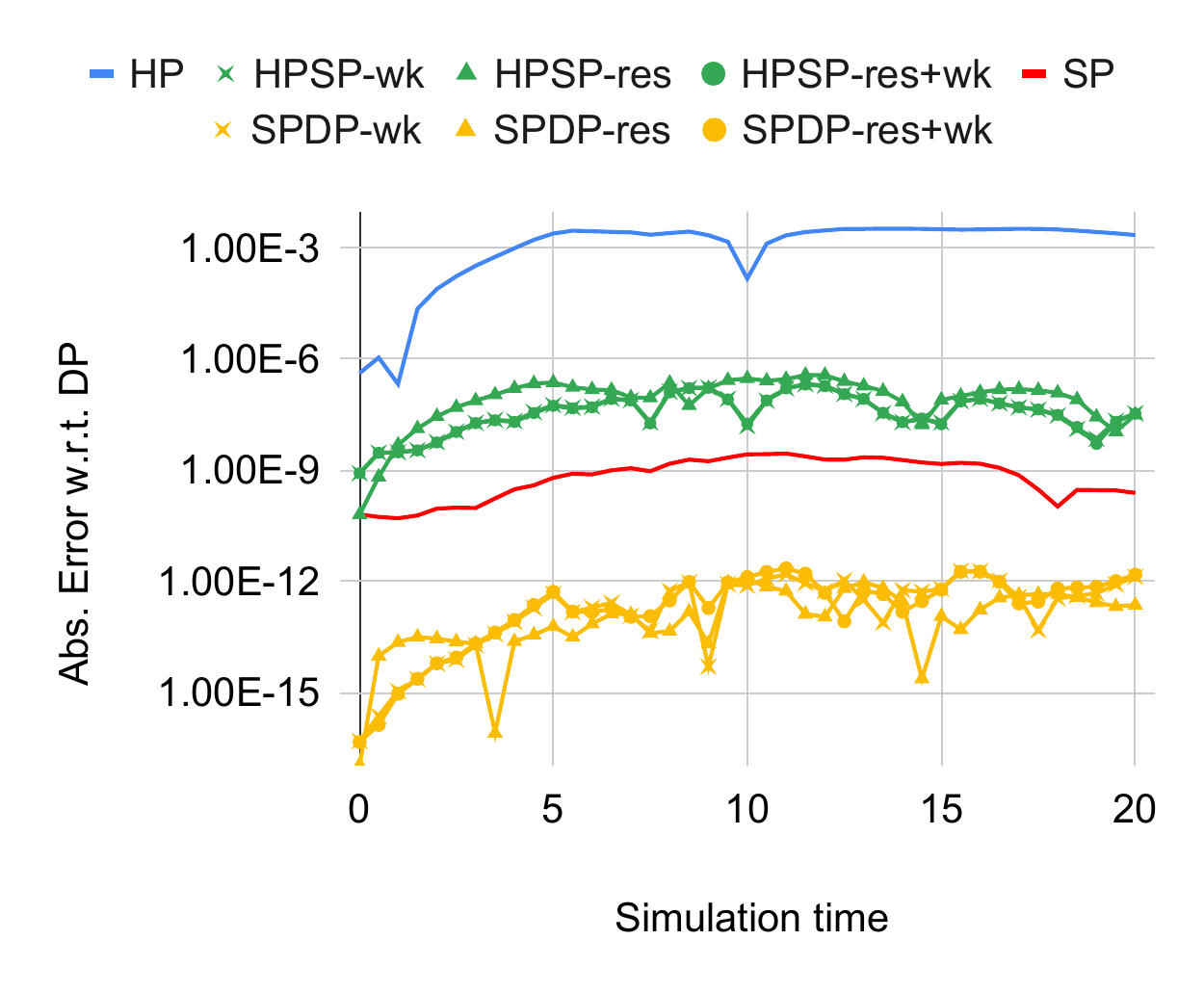}
        \caption{Absolute difference of dissipation, compared to double precision}
        \label{fig:TGsym_256c_dt0p0025_precision_diss_absdiff}
    \end{subfigure}
    \caption{Numerical accuracy of TGsym app using different precision levels. Mesh size $=256^3$, $M=0.5$, $Re=800$. The simulations were run for 8000 iterations using the default method.}
    \label{fig:TGsym_256c_dt0p0025_precision}
\end{figure}

Figure \ref{fig:TGsym_256c_dt0p0025_precision_diss} shows the effect of numerical precision on the behavior of the solenoidal dissipation rate as a function of time. Three standard versions are shown: HP using FP16 exclusively, SP using FP32, and DP using FP64. In addition, three kinds of mixed precision cases, namely `-wk', `-res', and `-res+wk', are presented between each pair of pure precisions. The `-wk' cases store only the work arrays in the lower precision, the `-res' cases store only the residual arrays in the lower precision, and the `-res+wk' cases store both the residual and work arrays in the lower precision. All test cases demonstrate overlapping results, with the exception of the FP16 version. The consistency observed among the other cases, with overlapping dissipation profiles, suggests comparable physical outcomes across these precision levels. Figure \ref{fig:TGsym_256c_dt0p0025_precision_diss_absdiff} presents a more detailed comparison and magnifies the differences by showing the absolute differences between each case and the highest precision run (FP64). The mixed-precision versions exhibit closer alignment with the higher precision of the two component precisions. Given the similarity in the observed error among the three mixed precision strategies, subsequent analyses will focus solely on the `-res+wk' strategy, as it offers the most significant computational speedup.

\begin{figure*}[htbp]
    \centering
    \begin{subfigure}[b]{0.49\linewidth}
        \centering
        \includegraphics[width=\linewidth]{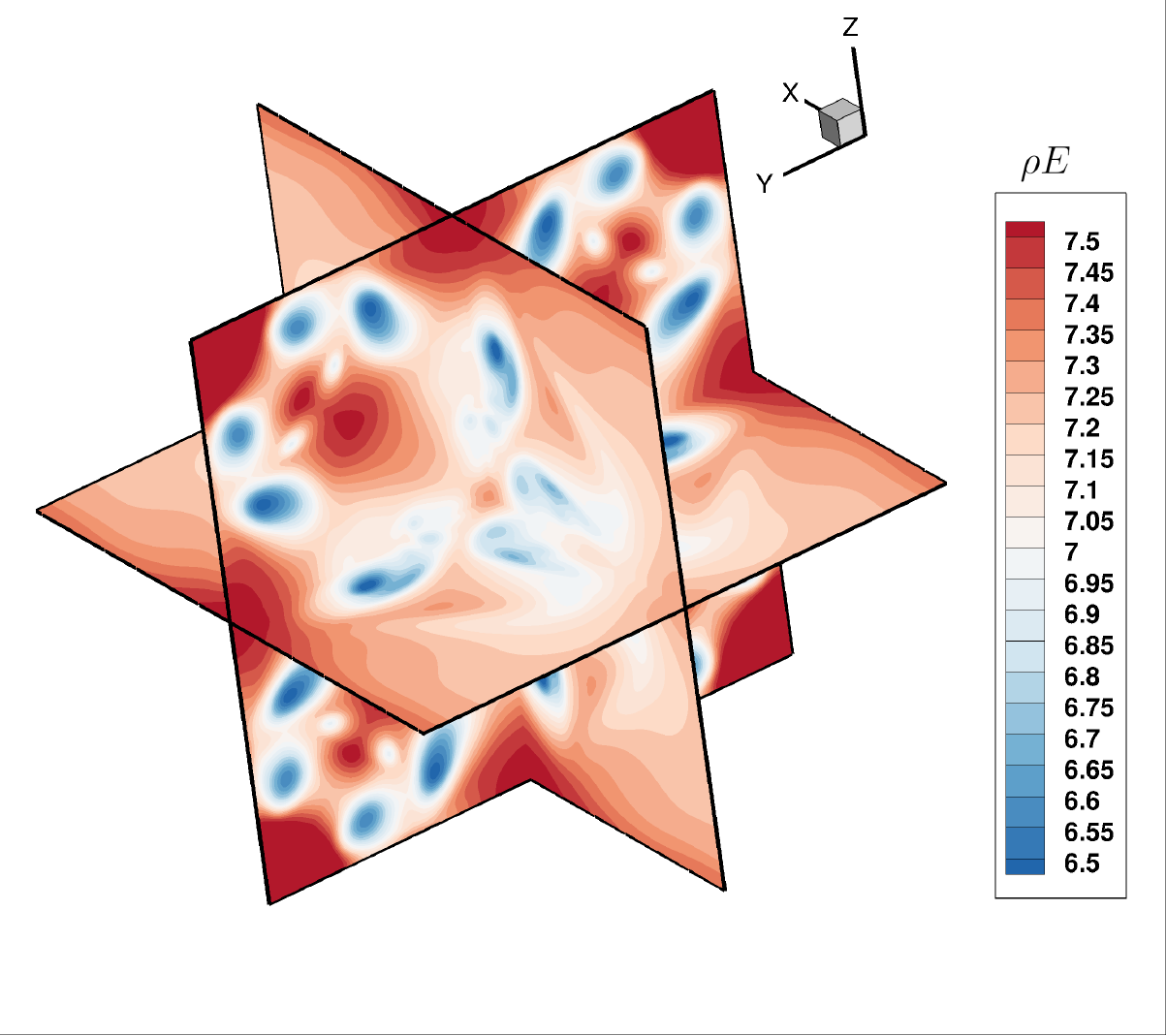}
        \caption{\textbf{SPDP mixed} TGV state at $t=10$.}
    \end{subfigure}
    \begin{subfigure}[b]{0.49\linewidth}
        \centering
        \includegraphics[width=\linewidth]{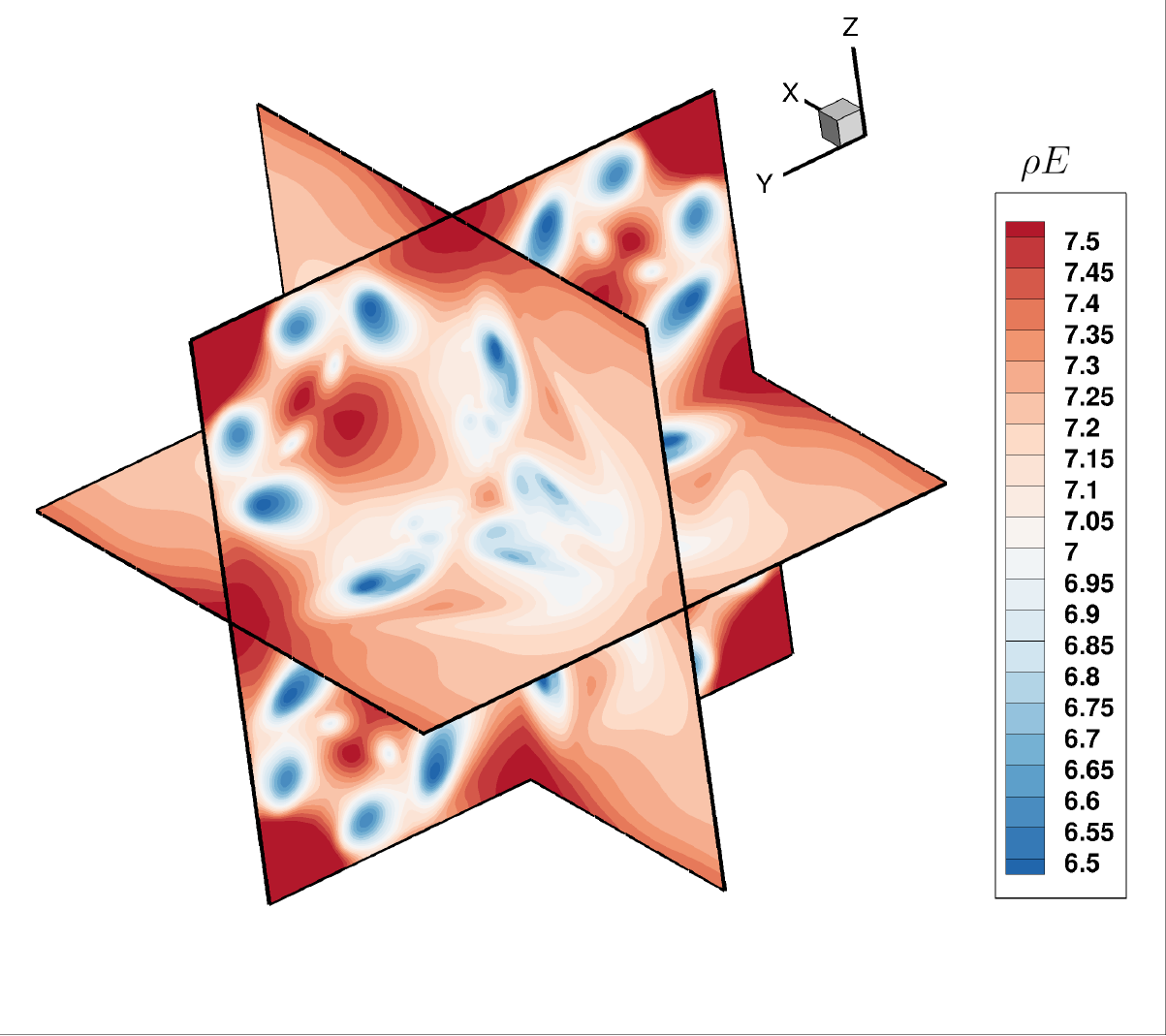}
        \caption{\textbf{SP} TGV state at $t=10$.}
    \end{subfigure} 
    \begin{subfigure}[b]{0.49\linewidth}
        \centering
        \includegraphics[width=\linewidth]{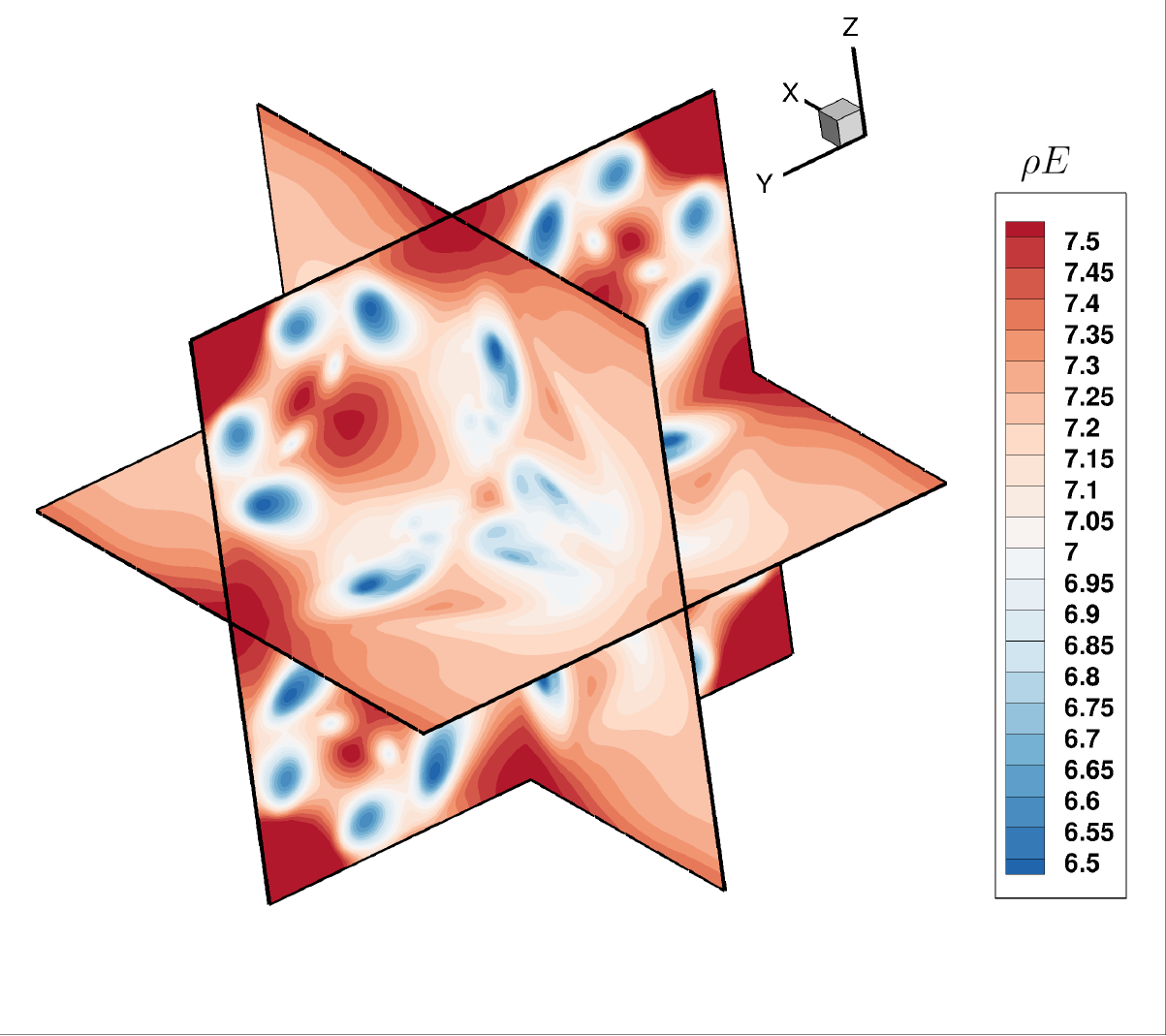}
        \caption{\textbf{HPSP mixed} TGV state at $t=10$.}
    \end{subfigure}     
    \begin{subfigure}[b]{0.49\linewidth}
        \centering
        \includegraphics[width=\linewidth]{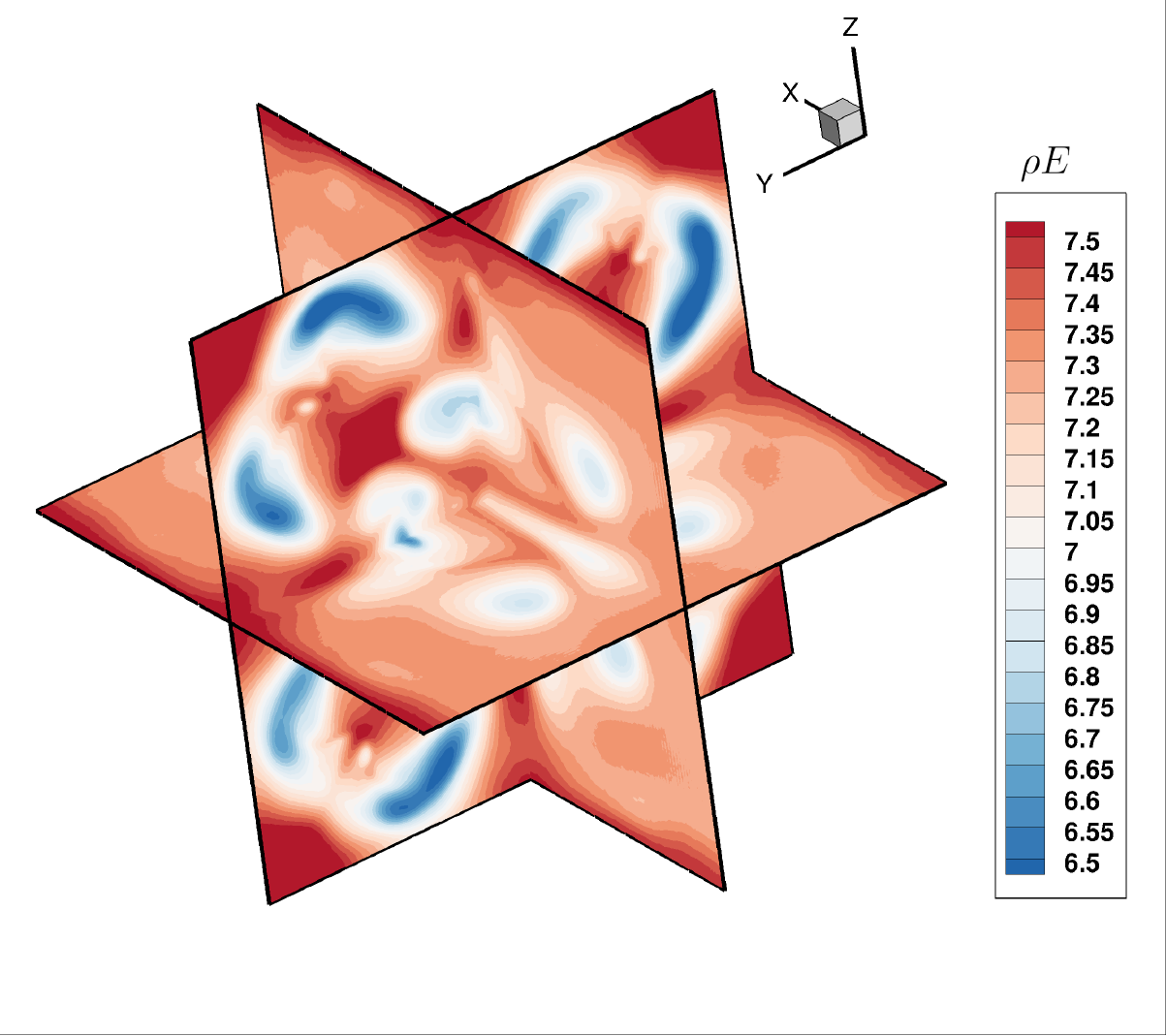}
        \caption{\textbf{HP} TGV state at $t=10$.}
    \end{subfigure}     
    \caption{Contours of $\rho E$ showing the TGV state at $t=10$, close to the peak of dissipation: (a) SPDP, (b) SP, (c) HPSP and (d) HP.}
    \label{fig:TGV_visualization1}
\end{figure*}

Figure \ref{fig:TGV_visualization1} shows the state of TGV at $t=10$ for various types of precision computations. It is clear that the flow state is qualitatively the same for lower precision and mixed precision computations all the way to the HPSP mixed setup, when compared to the most accurate DP results presented in Fig. \ref{fig:TGV_evolutions_t_10}. The results only deviate for the pure half-precision computations (HP), as can be noted from the last frame in the figure, where the vortex structures are more diffuse and sometimes in different locations. 

\begin{figure}[t]    
    \begin{subfigure}[b]{0.8\linewidth}
        \centering
        \includegraphics[width=\linewidth]{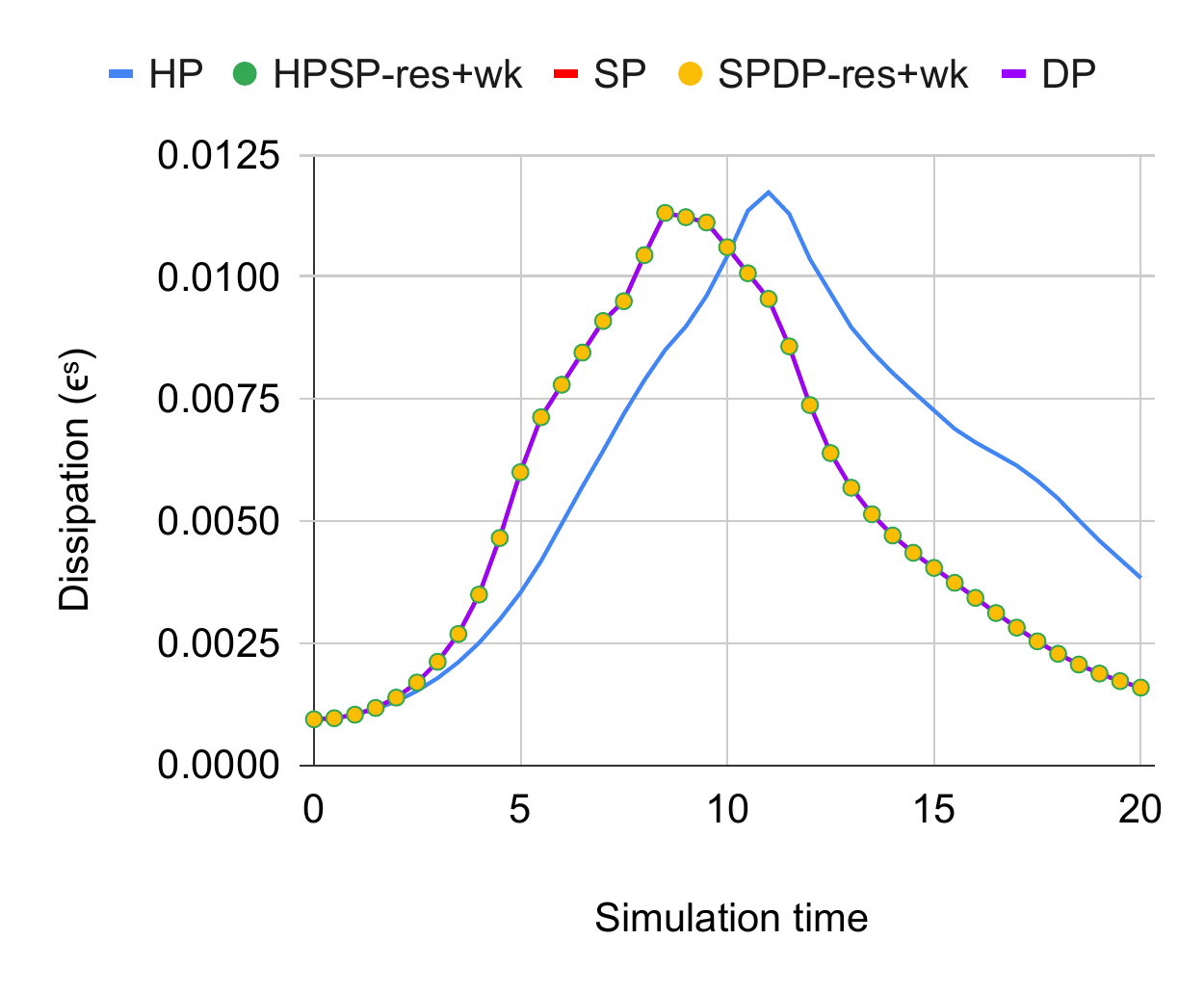}
        \caption{Dissipation}%\\\phantom{B}}
        \label{fig:TGsym_storesome_256c_dt0p0025_precision_diss}
    \end{subfigure}
    \hfill
    \begin{subfigure}[b]{0.8\linewidth}
        \centering
        \includegraphics[width=\linewidth]{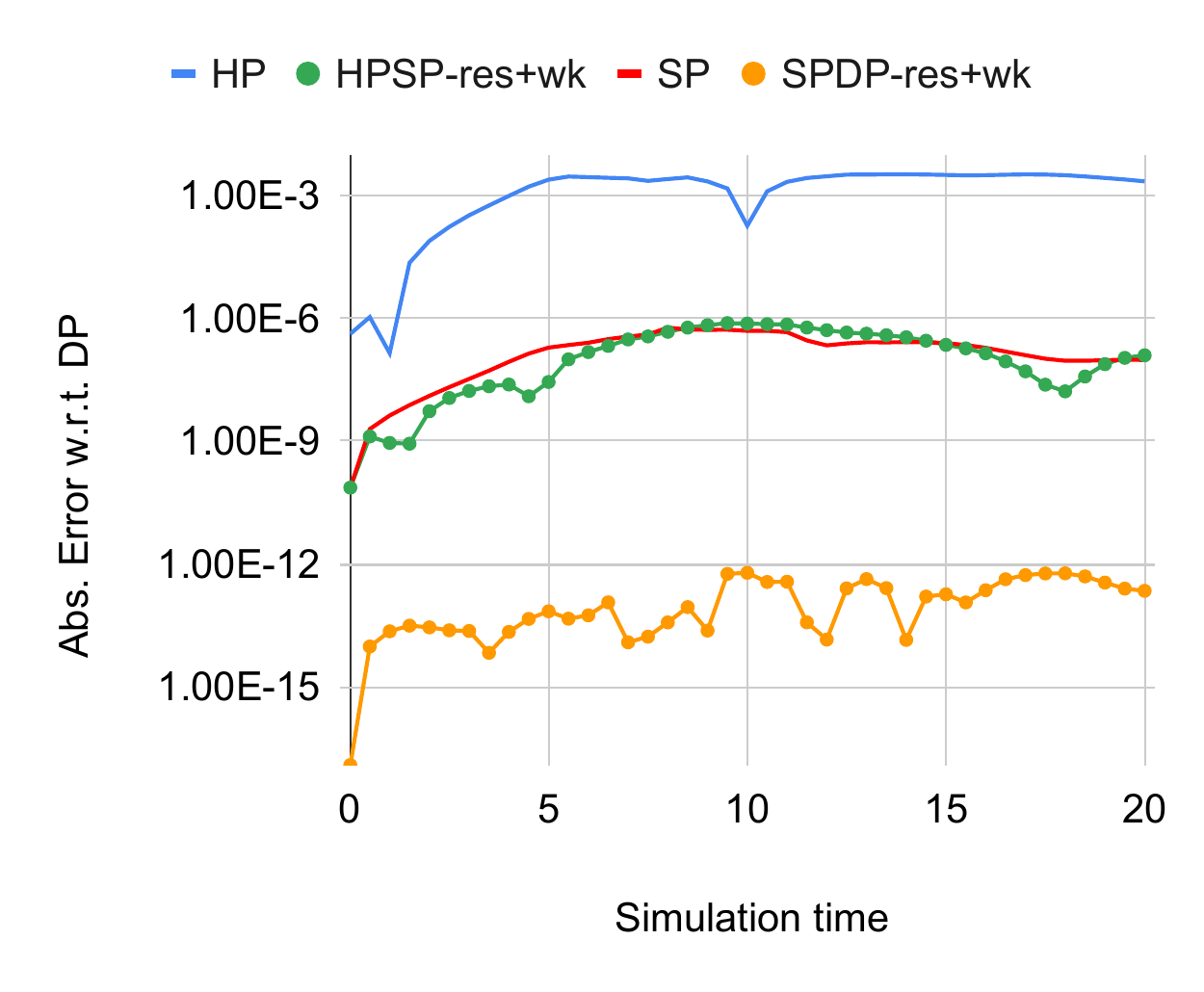}
        \caption{Absolute difference of dissipation, compared to double precision}
\label{fig:TGsym_storesome_256c_dt0p0025_precision_diss_absdiff}
    \end{subfigure}
    \caption{Numerical accuracy of TGsym app using different precision levels.  Mesh size $=256^3$, $M=0.5$, $Re=800$. The simulations were run for 8000 iterations using the Storesome method.}
    \label{fig:TGsym_storesome_256c_dt0p0025_precision}
\end{figure}

The accuracy of the Storesome method is presented in Figure \ref{fig:TGsym_storesome_256c_dt0p0025_precision} in the same format as Figure \ref{fig:TGsym_256c_dt0p0025_precision}. This method also demonstrates a high level of accuracy across different configurations, with the exception of the half-precision (FP16) run, which yielded inaccurate results. The overlapping results for all the other precision configurations indicate comparable physical outcomes. From Figure \ref{fig:TGsym_storesome_256c_dt0p0025_precision_diss_absdiff}, we can see how the levels of error varied among the mixed-precision runs. Notably, as the number of work arrays decreases significantly in the Storesome approach, a larger portion of the data is retained in higher precision. Consequently, a mixed half-single precision run exhibits a similar level of accuracy to that of a pure single precision run.

\begin{figure}[t]    
    \begin{subfigure}[b]{0.8\linewidth}
        \centering
        \includegraphics[width=\linewidth]{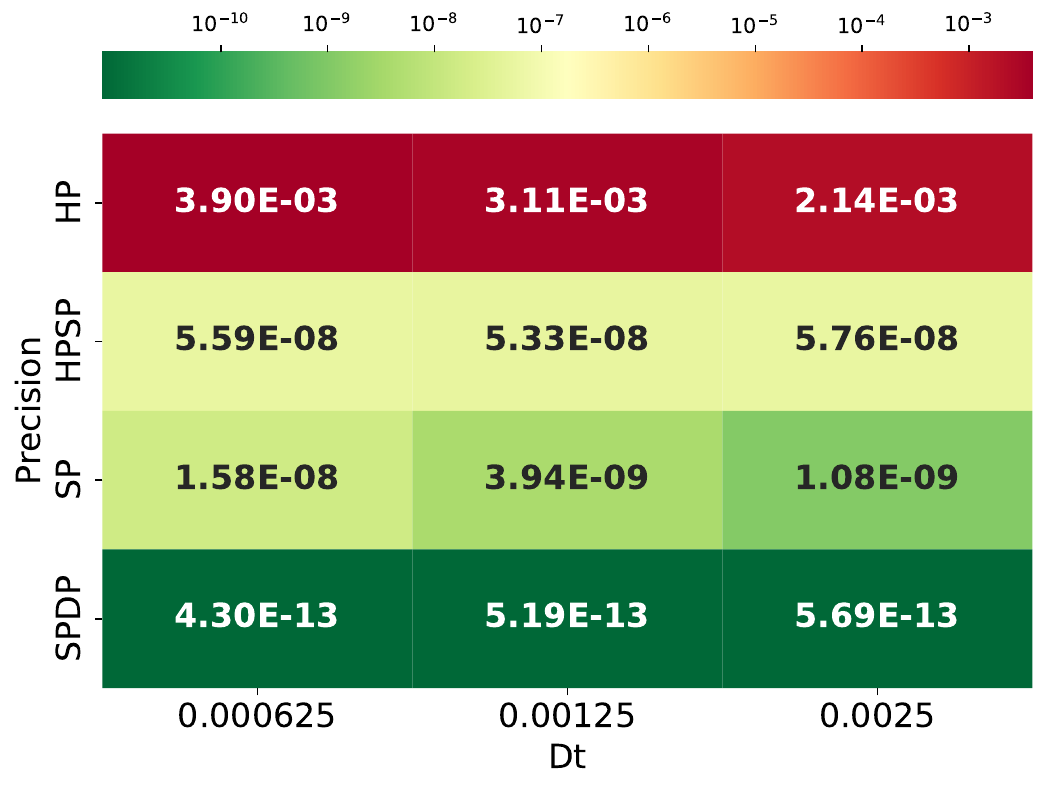}
        \caption{Effect of changing timestep. $M=0.5$, $Re=800$, $20/dt$ iterations.}
        \label{fig:heatmap_dt}
    \end{subfigure}
    \hfill
    \begin{subfigure}[b]{0.8\linewidth}
        \centering
        \includegraphics[width=\linewidth]{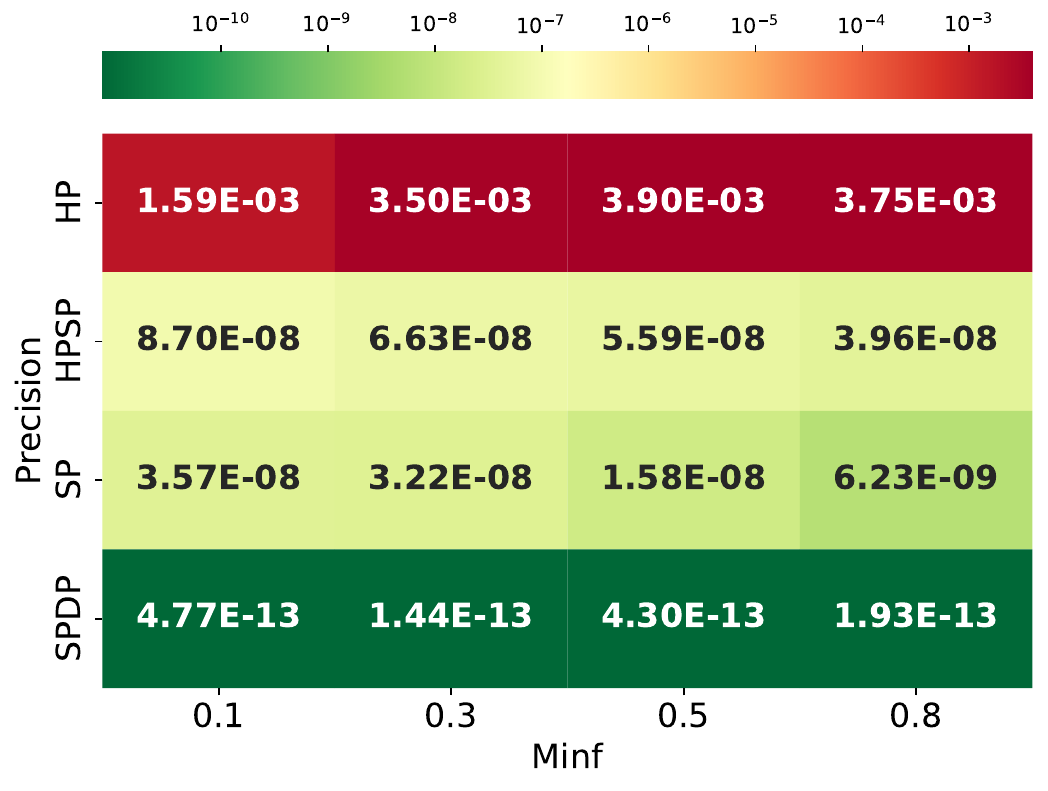}
        \caption{Effect of changing Mach number $M$. $Re=800$, 32000 iterations, $dt=0.000625$}
        \label{fig:heatmap_minf}
    \end{subfigure}
    \caption{The effect of changing parameters of the model on the numerical accuracy of TGsym app using different precision levels. Size=$256^3$, default method. Values are the average of the absolute differences against the DP run of the dissipation at every 0.5 stepsize of simulation time.}
    \label{fig:heatmaps}    
\end{figure}

Figure \ref{fig:heatmaps} illustrates the impact of varying the timestep and Mach number $M$ on the precision of the results. Small timesteps are required for low Mach number simulations and imply the additions of progressively smaller updates to the flow variables that are potentially sensitive to reduced precision. Here we separate out the effects of time step from those of Mach number. In Figure \ref{fig:heatmap_dt}, where the time step is varied while keeping the Mach number constant, it is evident that the reduced precision runs, where all variables are maintained at the same precision level (HP or SP), yield slightly less accurate results with smaller timesteps. However, when residuals and work arrays are set to lower precision, changes in timestep do not affect the results. This is because the $Q$ array is retained in higher precision, allowing for more accurate result storage. All observed errors remain within a factor of 10 of the smallest representable number for each specific precision. Figure \ref{fig:heatmap_minf} shows the effect of Mach number, while maintaining the same time step. While the HP case is incorrect for all Mach numbers, only a small increase in error is seen for the HPSP, SP, and SPDP cases as the Mach number is reduced.  Overall, both subplots indicate that mixed/reduced precision runs can be effectively utilized.

\begin{figure}[htbp]    
    \centering
    \includegraphics[width=.9\linewidth]{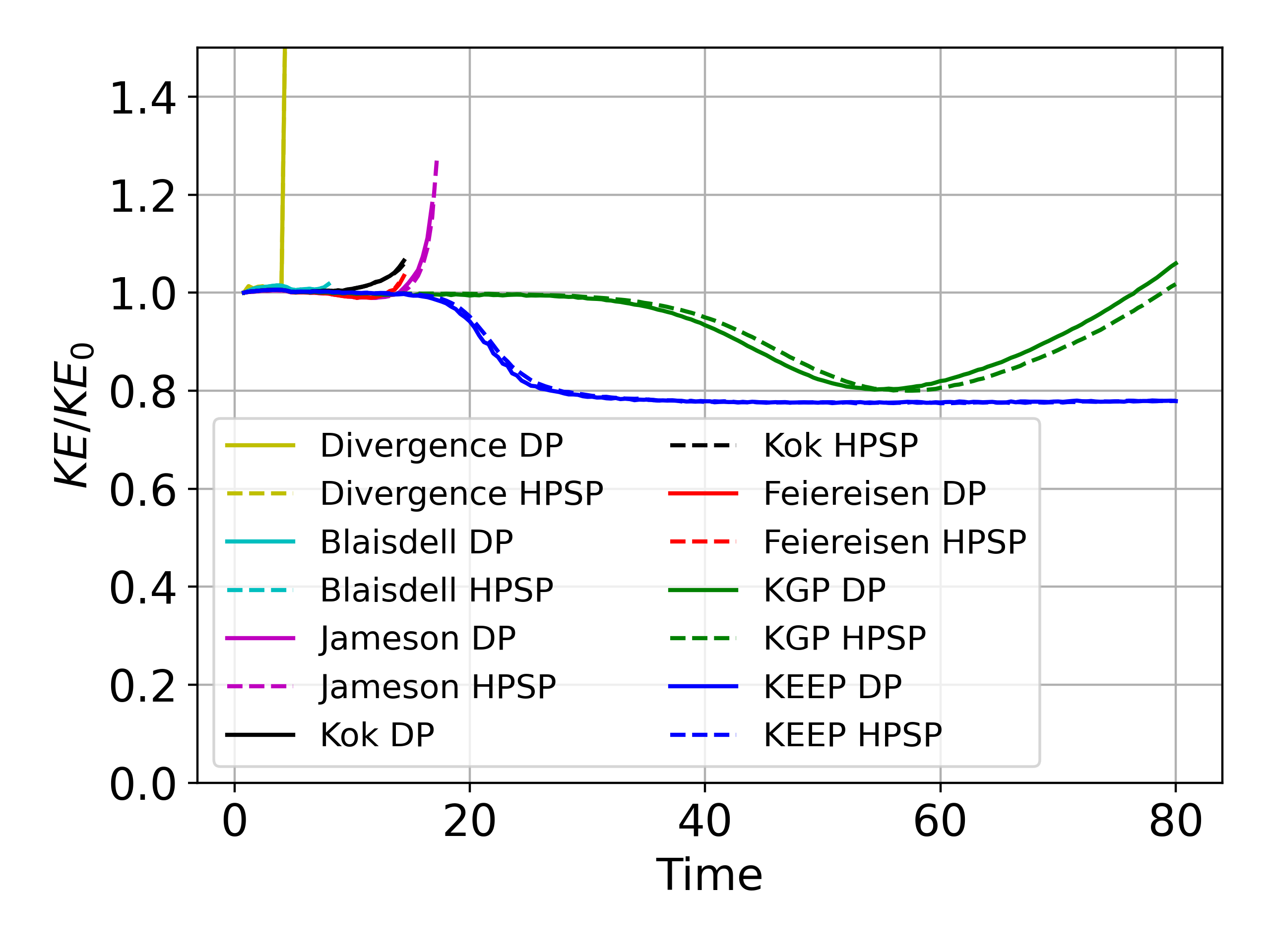}
    \caption{The effect of different split-forms on numerical accuracy of the inviscid Taylor-Green vortex application using DP and HPSP precision levels. $N=64^3$ grid points, $M_\infty=0.4$, $dt=0.004$, inviscid calculation, StoreSome method.}
    \label{fig:splitschemes}    
\end{figure}

We next consider a selection of the convective split-forms that are available in the OpenSBLI solver \cite{OpenSBLI_2024_CPC_V3} to improve simulation robustness. As we have already observed that half precision does not produce consistent results for the TGV application (Figure~\ref{fig:TGsym_storesome_256c_dt0p0025_precision}), we limit the discussion here only to the HPSP algorithm compared to the full double precision (DP) result. A full list of the equations for each of the split forms is given in the appendix of \cite{KUYA_KEEP_2018}. In order to isolate the relationship between precision and split-formulation of the equations, this comparison is performed as an inviscid TGV calculation with the diffusive terms omitted. We use a coarsened $N=64^3$ grid at $M_\infty = 0.4$ \cite{KUYA_KEEP_2018} to test the robustness of the split-forms at different numerical precisions.

Figure~\ref{fig:splitschemes} shows the time evolution of global kinetic energy normalised by its initial value. For a robust numerical scheme, the normalised kinetic energy for the inviscid case should stay at a value of one for as long as possible. The first thing to note in the figure is that, in the absence of physical viscosity in the inviscid limit considered here, several of the formulations (used for the convective terms of Navier-Stokes equations) diverge by showing exponential growth and eventually produce \texttt{NaN} when evaluated on a discrete mesh as non-linearities amplify aliasing errors. In particular, it can be observed that direct application of central differencing in the divergence form is unsuitable and the simulation rapidly diverges. Quadratic-split formulations by Blaisdell, Jameson, Kok, and Feiereisen (all listed in \cite{KUYA_KEEP_2018}) perform better, but still diverge within the standard simulation time used for Taylor-Green vortex problem (i.e. $t=20$). In contrast, the cubic-split Kennedy-Gruber-Pirozzoli (KGP) \citep{Coppola_CubicSplit_2019} and Kinetic Energy and Entropy Preserving (KEEP) \citep{KUYA_KEEP_2018} schemes are robust and are able to maintain numerical stability even in the inviscid limit on a coarse grid. The physical behaviour we observe between the different split formulations is in good agreement with previous studies of this problem \citep{KUYA_KEEP_2018}.

The effect of numerical precision is also shown on Figure~\ref{fig:splitschemes}, comparing double precision (solid line) with half-single precision (dashed line). Only very minor differences are observed between the full and mixed precision algorithms, with the largest differences being observed for the KGP scheme \citep{Coppola_CubicSplit_2019}. In each case, the trends between the split forms are consistent between double and half-single precision. This demonstrates that the improved numerical robustness of split formulations for the convective terms of the Navier-Stokes equations persists even when applying reduced/mixed numerical precision. The benefits of the mixed precision strategies are maintained when using other formulations of the equations, and are therefore not limited to only the baseline split-form.

To conclude this section, we highlight the findings regarding the accuracy of mixed and half precision formats in our simulations. Our evaluations demonstrate that while mixed precision configurations, such as half-single precision (HPSP), maintain acceptable accuracy levels, pure half precision (HP) fails to meet the required standards across various conditions. Importantly, these results are consistent regardless of the simulation parameters, such as Reynolds number and Mach number, as well as the choice of numerical schemes. This reinforces the notion that while mixed precision techniques can be effectively employed in computational fluid dynamics (CFD) applications, careful selection of precision is crucial to ensure reliable outcomes. As we move forward in our research, understanding these accuracy implications will be essential for optimizing performance without compromising the integrity of simulation results.

\subsection{Performance measurements}

In the previous sections, we demonstrated how the usage of lower precision affects accuracy at the application level. In this subsection, we will examine the changes in performance and quantify the trade-offs involved. Under normal circumstances, a complete TGV simulation (mesh size $=256^3$ $M=0.5$. $Re=800$, 8000 iterations, $\Delta t=0.0025$) using FP64 precision on a GPU takes approximately 825.68 seconds. In contrast, we observed that the simulation rearranged according to the Storesome method is significantly faster, completing in just 402.48 seconds while utilizing nearly half as much memory. This efficiency allows us to conduct more or even larger simulations within given hardware constraints while ensuring that the final results remain qualitatively comparable. For the performance measurements, we retained the mesh size of $256^3$ but ran the simulations for only 10\% of the total time steps (i.e. to time $t=2$), executing each case five times and reporting the average of these runs. We then plotted the compute time per iteration, as this metric remains consistent throughout a simulation.
Through our analysis of the representative TGV application, we illustrate what performance improvements users can expect in terms of both time and scale. For context, JAXA's aerofoil buffet applications run for three to four weeks using 120 Nvidia V100 GPUs~\cite{OpenSBLI_2024_CPC_V3,lusher2025highfidelity_JFM}. This highlights that running a substantial industrial application is not cheap; thus, any performance enhancement can be highly beneficial.

\begin{table}[ht]
    \centering
\begin{tabular}{l|rr||rr}
     & \multicolumn{2}{c}{Default}                                       & \multicolumn{2}{c}{Storesome}                                    \\ \hline
     & \multicolumn{1}{l}{runtime} & \multicolumn{1}{l}{speedup} & \multicolumn{1}{l}{runtime} & \multicolumn{1}{l}{speedup} \\ \hline
HP   & {42.43 ms}         & 2.43 $\times$                           & 18.67 ms                 & 2.69 $\times$                            \\
HPSP & {47.71 ms}         & 2.16 $\times$                           & 22.13 ms                 & 2.27 $\times$                            \\
SP   & {56.95 ms}         & 1.81 $\times$                           & 25.00 ms                 & 2.01 $\times$                            \\
SPDP & {74.02 ms}         & 1.39 $\times$                           & 37.60 ms                 & 1.34 $\times$                            \\
DP   & {103.21 ms}        & 1.00 $\times$                           & 50.31 ms                 & 1.00 $\times$                          
\end{tabular}
    \caption{Runtime per iteration and speedup compared to the double precision run time of the TGsym app are shown for the default and storesome generation methods. Mesh size=$256^3$, $M=0.5$, Re=800, 800 iterations. The measurements are performed on a single NVIDIA A100-SXM4-40GB GPU with an AMD EPYC™ 7763 (Milan) CPU.}
    \label{tab:TGsym256_speedups_CUDAGPU}
\end{table}

\begin{table}[ht]
    \color{red}
    \centering
\begin{tabular}{l|rr||rr}
     & \multicolumn{2}{c}{Default}                                       & \multicolumn{2}{c}{Storesome}                                    \\ \hline
     & \multicolumn{1}{l}{runtime} & \multicolumn{1}{l}{speedup} & \multicolumn{1}{l}{runtime} & \multicolumn{1}{l}{speedup} \\ \hline
HP   & {75.09 ms}         & 2.31 $\times$                           & 33.90 ms                 & 3.00 $\times$                            \\
HPSP & {90.23 ms}         & 1.92 $\times$                           & 48.65 ms                 & 2.09 $\times$                            \\
SP   & {108.91 ms}        & 1.59 $\times$                           & 53.84 ms                 & 1.89 $\times$                            \\
SPDP & {140.69 ms}        & 1.23 $\times$                           & 92.51 ms                 & 1.10 $\times$                            \\
DP   & {173.60 ms}        & 1.00 $\times$                           & 101.68 ms                & 1.00 $\times$                          
\end{tabular}
    \caption{\color{red} Runtime per iteration and speedup compared to the double precision run time of the TGsym app are shown for the default and storesome generation methods. Mesh size=$256^3$, $M=0.5$, Re=800, 800 iterations. The measurements are performed on a single AMD MI250X GPU with a 64-core AMD Trento CPU.}
    \label{tab:TGsym256_speedups_AMDGPU}
\end{table}

{\color{red}
The performance evaluation of the TGsym application, shown in Table~\ref{tab:TGsym256_speedups_CUDAGPU} and Table \ref{tab:TGsym256_speedups_AMDGPU}, reveals significant differences in runtime and speedup across various precision techniques on multiple GPU platforms. 
}
It is important to note that the Storesome method, by design, uses significantly fewer data arrays compared to the default method. This inherent efficiency of the Storesome approach contributes to its faster runtime performance by default. 

As we transition to lower precision methods, we observe a consistent increase in speedup. However, the approximate doubling of speedup ceases when using half precision (FP16). Currently, OPS does not support packed half precision types, such as \texttt{half2}, meaning that instruction throughput is still as if we were using FP32, and only the memory movement is reduced. The mixed precision configurations, specifically HPSP and SPDP, strike a balance between performance and accuracy, yielding speedups that fall between their respective pure precision counterparts (HP/SP and SP/DP).

{\color{red}
On the NVIDIA A100 GPU, the mixed precision strategies generally achieve notable speedups due to native support for half precision arithmetic. Similarly, the AMD MI250X GPU, which also features native FP16 hardware support, benefits from mixed precision configurations, offering enhanced performance by leveraging its FP16 capabilities.
}

\begin{table}[ht]
    \centering
\begin{tabular}{l|rr||rr}
     & \multicolumn{2}{c}{Default}                                       & \multicolumn{2}{c}{Storesome}                                    \\ \hline
     & \multicolumn{1}{l}{runtime} & \multicolumn{1}{l}{speedup} & \multicolumn{1}{l}{runtime} & \multicolumn{1}{l}{speedup} \\ \hline
HP   & {68.48 ms}         & 5.71 $\times$                           & 21.39 ms                 & 8.12 $\times$                            \\
HPSP & {105.22 ms}         & 3.71 $\times$                           & 47.31  ms                 & 3.67 $\times$                            \\
SP   & {185.23 ms}         & 2.11 $\times$                           & 65.05  ms                 & 2.67 $\times$                            \\
SPDP & {249.11 ms}         & 1.57 $\times$                           & 123.22 ms                 & 1.41 $\times$                            \\
DP   & {390.71 ms}         & 1.00 $\times$                           & 173.68 ms                 & 1.00 $\times$                          
\end{tabular}
    \caption{Runtime per iteration and speedup compared to the double precision run of the TGsym app using the default and Storesome generation methods. Mesh size=$256^3$, M=0.5, Re=800, 800 iterations. The measurements are performed on Intel Xeon Platinum 8592+ CPU} 
    \label{tab:TGsym256_speedups_intelCPU}
\end{table}

\begin{table}[ht]
    \centering
    \color{red}
\begin{tabular}{l|rr||rr}
     & \multicolumn{2}{c}{Default}                                       & \multicolumn{2}{c}{Storesome}                                    \\ \hline
     & \multicolumn{1}{l}{runtime} & \multicolumn{1}{l}{speedup} & \multicolumn{1}{l}{runtime} & \multicolumn{1}{l}{speedup} \\ \hline
HP   & { 78.78 ms}         & 4.48 $\times$                           & 47.76 ms                  & 2.80 $\times$                            \\
HPSP & {107.09 ms}         & 3.30 $\times$                           & 48.71  ms                 & 2.74 $\times$                            \\
SP   & {169.90 ms}         & 2.08 $\times$                           & 63.48  ms                 & 2.10 $\times$                            \\
SPDP & {228.69 ms}         & 1.54 $\times$                           & 105.17 ms                 & 1.27 $\times$                            \\
DP   & {353.32 ms}         & 1.00 $\times$                           & 133.49 ms                 & 1.00 $\times$                          
\end{tabular}
    \caption{\color{red}Runtime per iteration and speedup compared to the double precision run of the TGsym app using the default and Storesome generation methods. Mesh size=$256^3$, M=0.5, Re=800, 800 iterations. The measurements are performed on AMD EPYC Genoa (4th Gen) CPU} 
    \label{tab:TGsym256_speedups_AMDCPU}
\end{table}

\begin{table}[ht]
    \centering
    \color{red}
\begin{tabular}{l|rr||rr}
     & \multicolumn{2}{c}{Default}                                       & \multicolumn{2}{c}{Storesome}                                    \\ \hline
     & \multicolumn{1}{l}{runtime} & \multicolumn{1}{l}{speedup} & \multicolumn{1}{l}{runtime} & \multicolumn{1}{l}{speedup} \\ \hline
HP   & { 772.73 ms}         & 3.91 $\times$                           & 312.46 ms                  & 4.18 $\times$                            \\
HPSP & {1021.08 ms}         & 2.96 $\times$                           & 524.46 ms                 & 2.49 $\times$                            \\
SP   & {1536.12 ms}         & 1.97 $\times$                           & 616.89 ms                 & 2.12 $\times$                            \\
SPDP & {2013.77 ms}         & 1.50 $\times$                           & 1006.87 ms                 & 1.30 $\times$                            \\
DP   & {3020.79 ms}         & 1.00 $\times$                           & 1306.43 ms                 & 1.00 $\times$                          
\end{tabular}
    \caption{\color{red} Runtime per iteration and speedup compared to the double precision run of the TGsym app using the default and Storesome generation methods. Mesh size=$512^3$, M=0.5, Re=800, 800 iterations. The measurements are performed on Intel Xeon Platinum 8592+ CPU} 
    \label{tab:TGsym512_speedups_intelCPU}
\end{table}

Table \ref{tab:TGsym256_speedups_intelCPU} presents the performance of various methods on an Intel CPU platform. Unlike the GPU, this CPU utilizes fixed 512-bit wide vectors, allowing it to fully utilize these vectors with smaller data sizes thanks to compiler auto-vectorization. As a result, we observe the expected doubling of speedup even when employing FP16 precision. Additionally, the data indicates superlinear scaling, which occurs as more data fits into the cache memory. {\color{red} In contrast, Table \ref{tab:TGsym256_speedups_AMDCPU} shows that on an AMD CPU, which lacks FP16 vector optimizations, the performance does not improve significantly for HPSP and HP precision runs. Furthermore, Table \ref{tab:TGsym512_speedups_intelCPU} demonstrates that with a larger mesh size that exceeds cache capacity, the speedup approaches the expected 4x, thereby validating the superlinear speedup observed with the original mesh size.}

\begin{table}[ht]
    \centering
\begin{tabular}{l|rr||rr}
     & \multicolumn{2}{c}{Default}                                       & \multicolumn{2}{c}{Storesome}                                    \\ \hline
     & \multicolumn{1}{l}{Memory} & \multicolumn{1}{l}{Gain} & \multicolumn{1}{l}{Memory} & \multicolumn{1}{l}{gain} \\ \hline
HP   & {2.21 GB}         & 4.00 $\times$                           & 1.09 GB                 & 4.00 $\times$                            \\
HPSP & {2.72 GB}         & 3.25 $\times$                           & 1.60 GB                 & 2.72 $\times$                            \\
SP   & {4.41 GB}         & 2.00 $\times$                           & 2.17 GB                 & 2.00 $\times$                            \\
SPDP & {5.43 GB}         & 1.63 $\times$                           & 3.19 GB                 & 1.36 $\times$                            \\
DP   & {8.83 GB}        & 1.00 $\times$                           & 4.35 GB                 & 1.00 $\times$                          
\end{tabular}
    \caption{Memory used with the TGsym app and memory gain compared to the double precision run. Size=$256^3$.}
    \label{tab:memoryused}
\end{table}

Table \ref{tab:memoryused} presents the memory consumption associated with each configuration. All mixed precision cases fall between the memory requirements of their respective pure precision counterparts. The degree to which data arrays are cast to lower precision directly influences the overall memory savings. For instance, when fewer arrays are converted, as seen with the Storesome method, the resulting memory savings are less pronounced, as are the resulting speedups.
Overall, reducing an application's memory footprint enables the accommodation of larger problems within the same hardware constraints. For example, by halving the data size in an application that fully utilizes a GPU's memory, we can effectively double the overall mesh size. This capability is crucial for improving computational efficiency and expanding the scope of solvable problems in high-performance computing environments.

\begin{figure}[htbp]
    \centering
    \includegraphics[width=.9\linewidth]{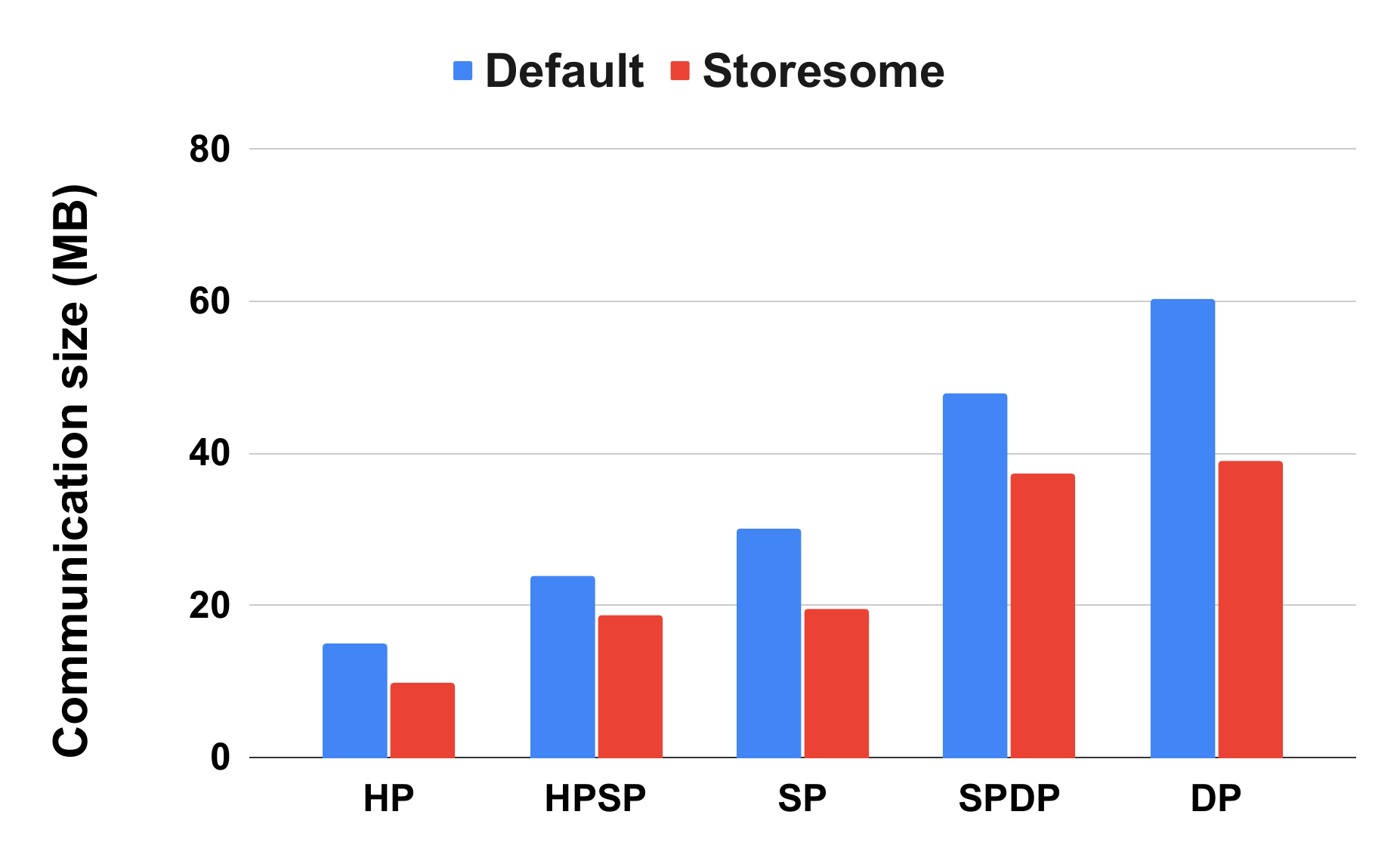}
    \caption{Average volume of MPI communications per process per iteration on the TGsym app, using 4 MPI processes. Size=$256^3$}
    \label{fig:MPIcomm}    
\end{figure}

Figure \ref{fig:MPIcomm} illustrates 
the communication requirements associated with utilizing multiple MPI processes. It is important to note that inter-node communication (usually through Infiniband) can be significantly slower than intra-node communication (usually through NVLink). Consequently, the reduction in data sizes has a pronounced effect on speedup when scaling across multiple devices. This figure also emphasizes the impact of selecting arrays to use in lower precision during mixed configurations. Since work arrays are communicated less frequently than other state arrays and the Storesome method utilizes fewer work arrays, the result is that mixed setups do not substantially reduce the MPI communication volume when employing the Storesome approach.

\begin{figure}[t]
    \centering
    \begin{subfigure}[b]{0.9\linewidth}
        \centering
        \includegraphics[width=\linewidth]{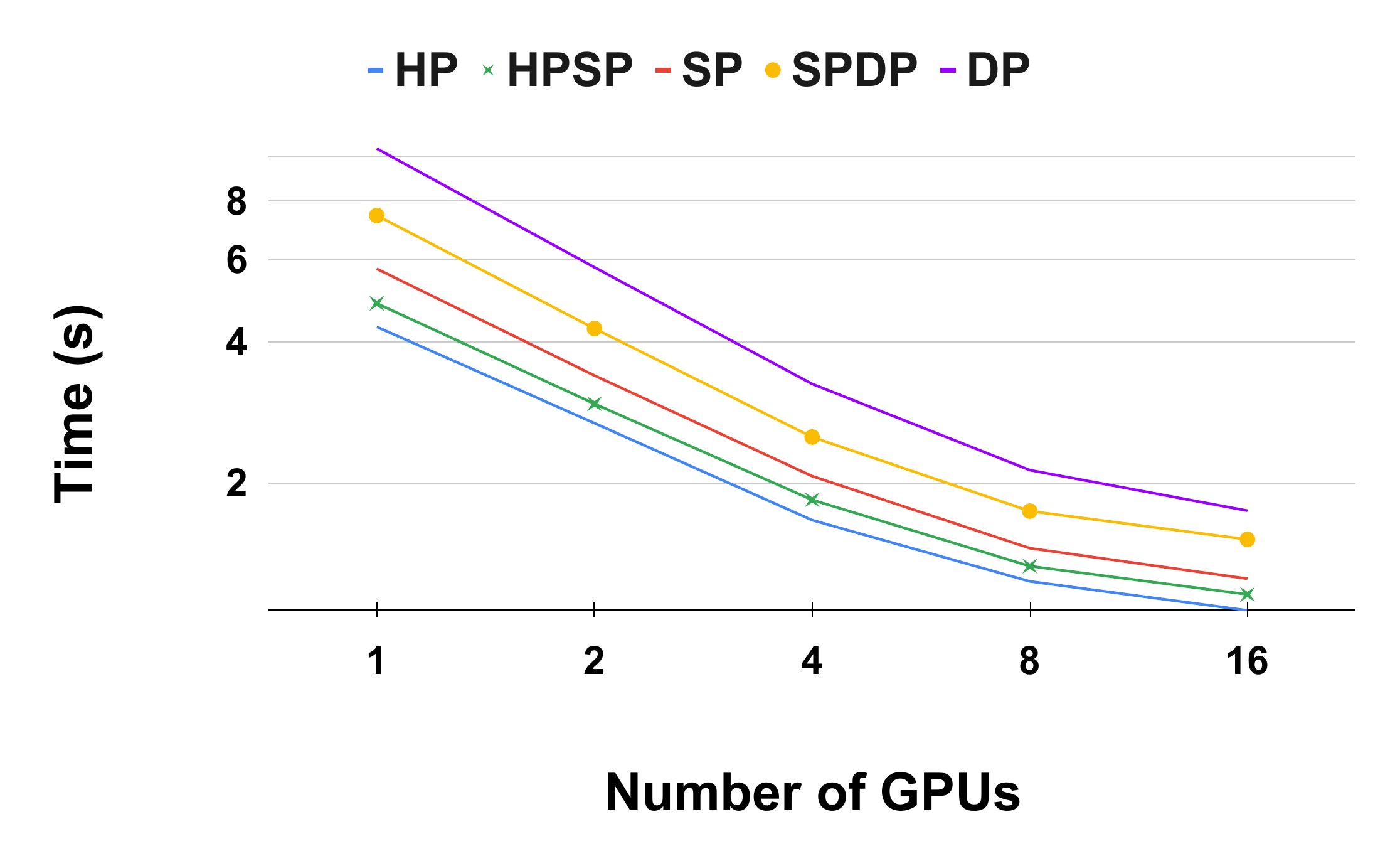}
        \caption{Strongscaling. Mesh size$=256^3$}
        \label{fig:TGsym_strongscaling}
    \end{subfigure}
    % \hfill
    \begin{subfigure}[b]{0.9\linewidth}
        \centering
        \includegraphics[width=\linewidth]{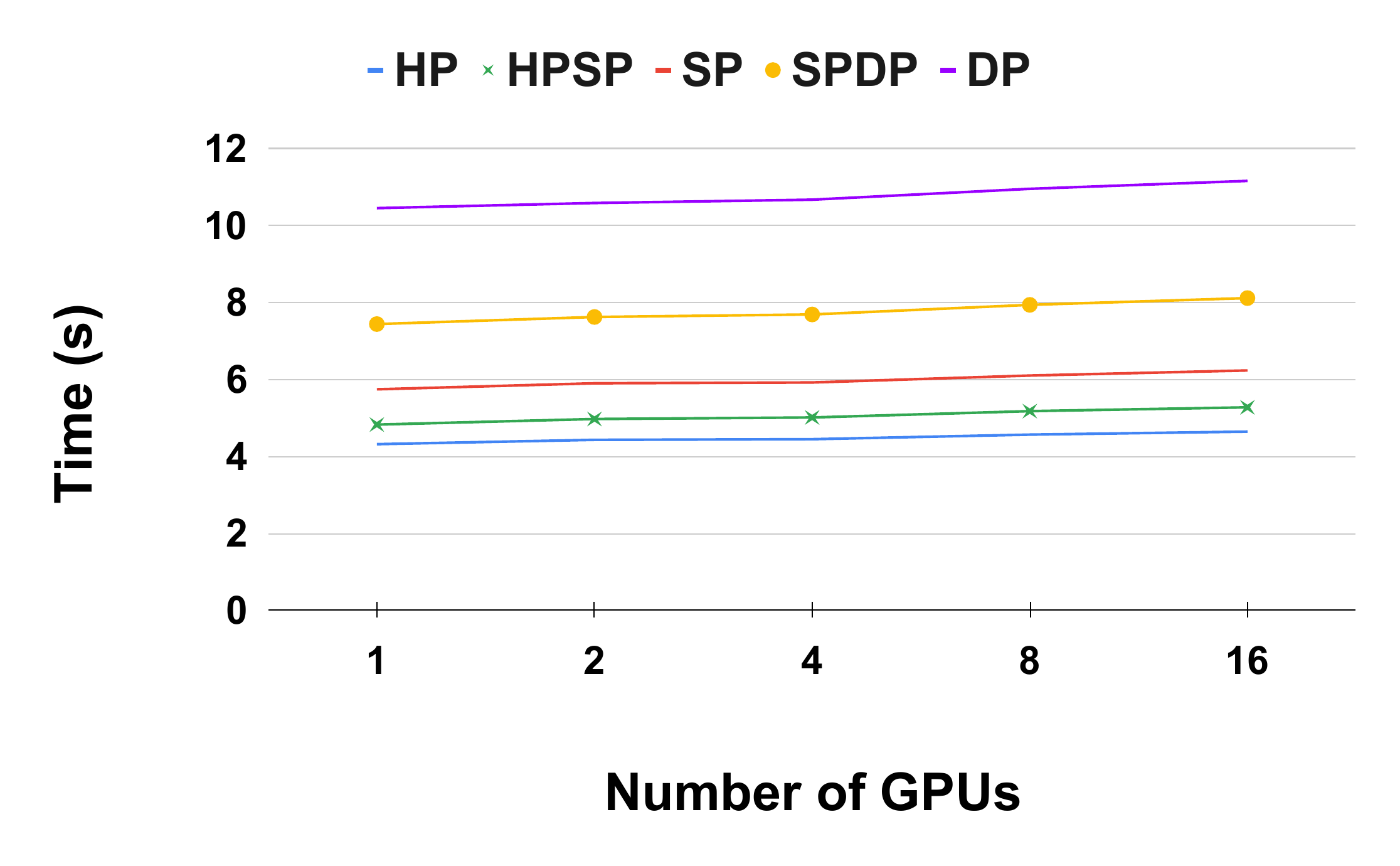}
        \caption{Weakscaling, Mesh size$=256^3$, multiplied by the number of processes in X direction.}
        \label{fig:TGsym_weakscaling}
    \end{subfigure}
     
    \caption{Strong- and weakscaling of the TGsym app with GPUdirect using different precision levels.  M=0.5, Re=800, 8000 iterations.}
    \label{fig:TGsym_scaling}    
\end{figure}

Figure \ref{fig:TGsym_scaling} illustrates both the strong and weak scaling performance of the TGsym application across different precision levels. As anticipated, all configurations demonstrate effective scaling behavior. The parallel efficiency results from the strong scaling measurements indicate effective scalability across all precision configurations, with efficiencies ranging from 90\% to 25\% as the number of processes increases. 
It is important to note that at higher process counts, message latency has a more significant impact on performance than message size. As the number of processes increases, the number of messages per process remains constant (and even slightly increases); however, the sizes of these messages decrease. This reduction in message size helps explain why double precision (DP) performance approaches that of single precision (SP) in these scenarios.
In contrast, the weak scaling measurements reveal consistently high efficiencies across all configurations, exceeding 92\%. This indicates that the TGsym application scales exceptionally well, irrespective of the precision level used. 

\section{Conclusions and Future Work}
\label{sec:concl}

In this paper, we have explored the implementation and effectiveness of mixed precision techniques in compressible turbulent flow simulations using explicit finite difference schemes. By extending both the OPS library and the OpenSBLI framework to support mixed precision arithmetic, we demonstrated that significant performance gains can be achieved without sacrificing numerical accuracy, provided that the precision is selected carefully. Our experiments using the Taylor-Green vortex benchmark revealed that while reduced precision formats such as single and mixed half-single precision yield acceptable results, pure half-precision computations suffer from unacceptable numerical inaccuracies.

The proposed mixed precision algorithm effectively balances performance and precision, allowing for improved computational efficiency, particularly in memory usage and communication overheads, which are crucial in multi-CPU and multi-GPU environments. Our analysis indicates that mixed precision approaches can provide substantial speedups, especially in larger-scale simulations, without compromising the integrity of the results.

Several avenues remain for further exploration:
\begin{itemize}
    \item Larger and More Complex Applications: To fully understand the potential of mixed precision techniques, {\color{red} future work will experiment with multi-block airfoils, with opportunities to use different precisions on different blocks, and shock-capturing schemes, in which some shock-capturing algorithms will need to be adjusted for reduced precision.} 
    %experiment with larger and more complex computational fluid dynamics (CFD) applications. 
    These experiments will also help assess the scalability and generalizability of our methods in real-world, high-performance computing (HPC) environments. 
    \item Improving Half-Precision Performance: Since pure half-precision (FP16) simulations did not yield acceptable accuracy in our current tests, we did not prioritize optimizing FP16 performance in this work. However, there may be specific simulations where FP16 can produce satisfactory results, particularly on GPU architectures. To support these scenarios, future work will focus on enhancing OPS to define global constants in half-precision and to utilize packed types, allowing multiple half-precision calculations within a single warp. These optimizations could unlock further performance gains on hardware optimized for FP16 computations, such as modern GPUs.
    \item {\color{red} As well as the possibility to treat different terms (for example viscous terms) with different precision, as noted in the introduction, it is also interesting to consider whether the precision can be adjusted in an adaptive way, for example in an analogous manner to adaptive mesh refinement, where the mesh is refined based on local error criteria.}
\end{itemize}

By addressing these areas, we aim to further optimize the trade-offs between computational efficiency and numerical accuracy, thereby pushing the boundaries of what is possible with mixed precision in large-scale turbulent simulations.

\section*{Acknowledgements}

This research was supported by the National Research, Development and Innovation Fund
of Hungary (FK 145931), under the FK\_23 funding scheme. We acknowledge KIF\"U (Governmental Agency for IT Development, Hungary, \url{https://ror.org/01s0v4q65}) for awarding us access to the Komondor HPC facility based in Hungary. NDS and PKS acknowledge the support of EPSRC under grant EP/W026686/1. DJL and the development of OpenSBLI V3.0 were supported by the Japan Society for the Promotion of Science (JSPS) under grant 22F22059. We are grateful for the support of the OneAPI Innovator program, and the advice and assistance of Rupak Roy, Omar Toral, Sri Ramkrishna at Intel in particular. 

\section*{Declaration of generative AI and AI-assisted technologies in the writing process.}
During the preparation of this work, the author(s) utilized both the Perplexity AI tool and ChatGPT to assist primarily in writing and phrasing, enhancing the overall flow of the text for better readability. After employing these tools, the author(s) reviewed and edited the material as necessary and take(s) full responsibility for the content of the published article.
% \appendix
% \section{Ex, ample Appendix Section}


 \bibliographystyle{elsarticle-num} 
 \bibliography{bibliography}

\end{document}